\documentclass[10pt,aps,prd,amsmath,amssymb,superscriptaddress,onecolumn,nofootinbib,showpacs,preprintnumbers,amsfonts, notitlepage,longbibliography]{revtex4-1}
\usepackage[utf8]{inputenc}

\usepackage[usenames,dvipsnames,svgnames,table]{xcolor} 
\usepackage{graphicx}% Include figure files
\usepackage{dcolumn}% Align table columns on decimal point
\usepackage{bm}% bold math
\usepackage{hyperref}% add hypertext capabilities
\hypersetup{ 
    pdfnewwindow=true,      % links in new window
    colorlinks=true,       % false: boxed links; true: colored links
    allcolors=[RGB]{65 125 10} % JLAB green
} 
\usepackage[capitalize]{cleveref}
\usepackage{dsfont}
\usepackage{placeins}
\usepackage{todonotes}
\usepackage[utf8]{inputenc}
\usepackage{gensymb}
\usepackage{multirow}
\usepackage{amsmath}
\usepackage{amsfonts}
\usepackage{amssymb}
\usepackage{enumitem,amssymb}
\usepackage{array}   % for \newcolumntype macro
\newcolumntype{L}{>{$}l<{$}} % math-mode version of "l" column type
\newcolumntype{R}{>{$}r<{$}}
\newcolumntype{C}{>{$}c<{$}}
\usepackage{xspace}
\usepackage{braket}
\usepackage{units}
\usepackage{slashed}
\usepackage[normalem]{ulem}
\usepackage[format=plain,justification=RaggedRight,singlelinecheck=false]{caption}
\usepackage[format=plain,justification=centering,singlelinecheck=false]{subcaption}
\usepackage{tabularx}
\usepackage{blindtext}

\graphicspath{{figures/}}
\allowdisplaybreaks

\newcommand{\rrvec}[0]{{\overrightarrow{r_n}}}
\newcommand{\lrvec}[0]{{\overleftarrow{r_n}}}

\newcommand{\Ac}[0]{{\mathcal{A}}}
\newcommand{\Bc}[0]{{\mathcal{B}}}
\newcommand{\Cc}[0]{{\mathcal{C}}} 
\newcommand{\Fc}[0]{{\mathcal{F}}}

\newcommand{\Hc}[0]{{\mathcal{H}}}

\newcommand{\Jc}[0]{{\mathcal{J}}}
\newcommand{\Mc}[0]{{\mathcal{M}}} 
\newcommand{\Rc}[0]{{\mathcal{R}}} 
\newcommand{\Tc}[0]{{\mathcal{T}}}
\newcommand{\Wc}[0]{{\mathcal{W}}}
\newcommand{\Yc}[0]{{\mathcal{Y}}}

\newcommand{\df}[0]{\mathrm{df}}
\newcommand{\diff}[0]{\mathrm{d}}
\newcommand{\nn}[0]{\nonumber}

\newcommand{\ggpipi}{\ensuremath{\gamma^{\star} \gamma^{\star} \to \pi \pi}\xspace}

\newcommand{\jlab}{
	Thomas Jefferson National Accelerator Facility,
	12000 Jefferson Avenue, 
	Newport News, VA 23606, USA
}
\newcommand{\odu}{
	Department of Physics,
	Old Dominion University,
	Norfolk, Virginia 23529, USA
}
\newcommand{\wm}{
	Department of Physics, 
	College of William and Mary, 
	Williamsburg, VA 23187, USA
}

\begin{document}

%%%%%%%%%%%%%%%%%%%%%%%%%%%%%%%%%%%%
%	Title
%%%%%%%%%%%%%%%%%%%%%%%%%%%%%%%%%%%%
\title{Prospects for \ggpipi via lattice QCD}

%%%%%%%%%%%%%%%%%%%%%%%%%%%%%%%%%%%%
%	Author list
%%%%%%%%%%%%%%%%%%%%%%%%%%%%%%%%%%%%

%%%%%%%%%%
\author{Ra\'ul~A.~Brice\~no}
\email[e-mail: ]{rbriceno@jlab.org}
\affiliation{\jlab}
\affiliation{\odu}
%%%%%%%%%%

%%%%%%%%%%
\author{Andrew~W.~Jackura}
\email[e-mail: ]{ajackura@jlab.org}
\affiliation{\jlab}
\affiliation{\odu}
%%%%%%%%%%

%%%%%%%%%%
\author{Arkaitz~Rodas}
\email[e-mail: ]{arodas@jlab.org}
\affiliation{\jlab}
\affiliation{\wm}
%%%%%%%%%%

%%%%%%%%%%
\author{Juan~V.~Guerrero}
\email[e-mail: ]{juanvg@jlab.org}
\affiliation{\jlab}
%%%%%%%%%%

%%%%%%%%%%%%%%%%%%%%%%%%%%%%%%%%%%%%
%	Abstract
%%%%%%%%%%%%%%%%%%%%%%%%%%%%%%%%%%%%
\begin{abstract}
The \ggpipi scattering amplitude plays a key role in a wide range of phenomena, including understanding the inner structure of scalar resonances as well as constraining the hadronic contributions to the anomalous magnetic moment of the muon. 
In this work, we explain how the infinite-volume Minkowski amplitude can be constrained from finite-volume Euclidean correlation functions. The relationship between the finite-volume Euclidean correlation functions and the desired amplitude holds up to energies where $3\pi$ states can go on shell, and is exact up to exponentially small corrections that scale like $\mathcal{O}(e^{-m_\pi L})$, where $L$ is the spatial extent of the cubic volume and $m_\pi$ is the pion mass. In order to implement this formalism and remove all power-law finite volume errors, it is necessary to first obtain $\pi \pi \to \pi\pi$, $\pi \gamma^\star \to \pi$, $\gamma^\star \to\pi\pi$, and $\pi\pi\gamma^\star \to\pi\pi$ amplitudes; all of which can be determined via lattice quantum chromodynamic calculations. 

\end{abstract}
\maketitle
\preprint{JLAB-THY-22-3727}

%%%%%%%%%%%%%%%%%%%%%%%%%%%%%%%%%%%%%%%%%%%%%%%%%%%%%%%%%%%%%%%%%%%%%%%%
%	Section :: Introduction
%%%%%%%%%%%%%%%%%%%%%%%%%%%%%%%%%%%%%%%%%%%%%%%%%%%%%%%%%%%%%%%%%%%%%%%%
\section{Introduction}
\label{sec:intro}

Several outstanding puzzles within the Standard Model of Particle Physics involve electroweak interactions of low-energy nuclear systems. One of the more pressing issues is the discrepancy between theoretical predictions and experimental measurements of the anomalous magnetic moment of the muon~\cite{Aoyama:2020ynm}, which is presently estimated to be at the $3\!-\!4\,\sigma$ deviation level~\cite{Muong-2:2021ojo}. If this tension persists, it would serve as indirect evidence of physics beyond the Standard Model. Non-perturbative effects from Quantum Chromodynamics (QCD) presently dominate the theoretical uncertainty, in particular, the QCD contributions to the hadronic vacuum polarization (HVP) and the hadronic light-by-light (HLbL) processes~\cite{Pauk:2014rfa,Colangelo:2014dfa,Colangelo:2014pva,Colangelo:2015ama,Nyffeler:2017ohp, Colangelo:2017fiz,Colangelo:2017fiz,Colangelo:2017qdm,Colangelo:2018mtw,Hoferichter:2018kwz,Hoferichter:2018dmo}. Estimating the size of these contributions from low-energy QCD is challenging, however, advancements in both phenomenology and theory have made substantial progress in decreasing their uncertainties. In the case of the HLbL tensor, dispersive representations are used to write the HLbL amplitude in terms of hadronic matrix elements such as $\gamma^{\star}\gamma^{\star} \to \pi^0$, $\eta$, $\eta'$, $\pi\pi$, $K\bar{K}$, \ldots, etc, which are in turn constrained from data-driven analyses~\cite{Danilkin:2017lyn,Danilkin:2018qfn,Danilkin:2019opj,Hoferichter:2019nlq, Dai:2014zta}.~\footnote{For complimentary lattice QCD efforts aimed at directly calculating the full HLbL contribution to the anomalous magnetic moment of the muon, we point the reader to Refs.~\cite{Chao:2021tvp, Chao:2020kwq, Blum:2019ugy}.}

In general, these two-photon processes can be obtained from the evaluation from matrix elements of time-separated products of electromagnetic currents, $\Jc^\mu$, overlapping between the QCD vacuum and the desired final state, $\sim \bra{\rm{out}}  \Jc^{\mu}(t)  \Jc^{\nu}(0)  \ket{\Omega} \,$. Accessing these classes of matrix elements directly from QCD requires a non-perturbative approach in order to correctly capture the low-energy physics of hadrons. Lattice QCD is a rigorous and controlled approximation to QCD where correlation functions are estimated using Monte Carlo techniques in a discrete Euclidean spacetime, confined in a finite volume. In recent years, there has been tremendous progress from the lattice QCD community, particularly in constraining increasingly complicated low-energy hadronic processes. This progress is largely due to three key components: (a) access to powerful high-performance computing resources, (b) development of sophisticated algorithms, and (c) development of non-perturbative formalisms that connect quantities directly accessed via lattice QCD and the desired scattering observables. The work presented here falls in the third category, enabling us to extract information on \ggpipi from lattice QCD calculations. 

Before discussing the conceptual challenges addressed in this work that would pave the way towards future determinations of the \ggpipi amplitude from lattice QCD, it is worth remarking one additional motivation for this amplitude. One of the earliest non-perturbative predictions of QCD is the presence of glueball states in the spectrum of the theory. In a pure Yang-Mills theory, these are hadronic states composed of bound gluons with no sea quarks. Several predictions have been performed in quenched calculations of the theory~\cite{Patel:1986vv,Albanese:1987ds,Michael:1988jr,Bali:1993fb,Sexton:1995kd,Morningstar:1999rf,Chen:2005mg,Athenodorou:2020ani}. Unfortunately, given the mixing between gluons and quark-anti-quark pairs in QCD, the identification of glueball states in the unquenched theory has proven to be an outstanding challenge, and to this day a ``\emph{smoking gun}" that is not model-dependent is missing. Given the fact that glueballs would necessarily be neutral states, one can expect that sensible quantitative measurements of glueballs include their radiative transitions. In particular, a large glueball component is expected to produce a small $\gamma \gamma$ coupling, as photons do not couple directly to gluons. The lowest-lying glueball candidate is expected to have $J^{PC} = 0^{++}$ quantum numbers and a mass of around 1.5 to 2 GeV. It couples to a myriad of final states, including $\pi \pi$, mixing with the inelastic $f_0(1370),\,f_0(1500)$, and $f_0(1710)$ resonances to a varying degree. Furthermore, in scattering processes, their  lineshapes overlap, which makes the glueball identification very challenging~\cite{Chanowitz:1980gu,Amsler:1995td,Amsler:1995tu,Lee:1999kv,Giacosa:2005zt,Giacosa:2005qr,Albaladejo:2008qa,Ochs:2013gi,Janowski:2014ppa,Pelaez:2019eqa,Rodas:2021tyb,Pelaez:2022qby}. As a result, the  \ggpipi amplitude can help provide some constraints on the inner structure of some glueball candidates.~\footnote{For a review on the role of two-photon couplings in elucidating the nature of light hadrons, we point the reader to Ref.~\cite{Pennington:2007yt} and references therein.}

Having motivated the desired amplitude, we can now discuss why this is challenging to access via lattice QCD. To have confidence in the determination of the \ggpipi amplitude, the QCD contribution must be treated non-perturbatively. Meanwhile, it is sufficient to treat the quantum electrodynamic (QED) contribution perturbatively using two electromagnetic currents inserted at arbitrary time separations. 
Directly computing such a process proves to be na\"ively impossible in lattice QCD because calculations are necessarily done in a finite, Euclidean spacetime. The fact that the space is finite, results in the absence of asymptotic states, including the asymptotic $\pi\pi$ state. While the Euclidean nature of the calculation generally prohibits the access of Minkowski matrix elements that explicitly depend on time, necessary for the product of previously discussed currents. However, here we discuss an indirect path towards accessing such an amplitude using Euclidean correlators in a finite volume.

    This work builds from an extensive program aimed at determining purely hadronic as well electroweak amplitudes via lattice QCD. Here we review the key work that makes this possible and provides evidence that the formalism is sufficiently mature to consider such a complex reaction as \ggpipi.  
    As is discussed in great detail in the remainder of this work, the formalism presented here requires inputs from a variety of scattering observables. The simplest of which are the purely-hadronic two-body scattering amplitudes. These amplitudes are not directly accessible in a finite spacetime. Instead, one can construct exact relations between the finite-volume spectrum, which is accessible via lattice QCD, and infinite-volume scattering amplitudes. This observation was first made by L\"uscher for non-perturbative theories involving identical bosons at rest~\cite{Luscher:1986pf}, and it has since been generalized to arbitrarily complicated two-body system~\cite{Rummukainen:1995vs,Kim:2005gf,Fu:2011xz,He:2005ey,Lage:2009zv,Bernard:2010fp,Briceno:2012yi,Hansen:2012tf,Feng:2004ua,Gockeler:2012yj,Briceno:2014oea,2012PhRvD..85k4507L}. This has allowed for numerous successful numerical calculations of scattering amplitudes~\cite{Dudek:2014qha,Guo:2018zss,Alexandrou:2017mpi,Prelovsek:2020eiw,Brett:2018jqw,Andersen:2017una,Wilson:2014cna,Briceno:2016mjc,Wilson:2019wfr,Wilson:2015dqa,Briceno:2017qmb,Gayer:2021xzv, Dudek:2016cru,  Moir:2016srx, Woss:2019hse, Woss:2020ayi,Rendon:2020rtw,Silvi:2021uya}. In particular, all that will be needed from this formalism are the $\pi\pi\to\pi\pi$ partial-wave amplitudes. 
    
The second class of  quantities needed are electromagnetic amplitudes involving a single current insertion coupling to $\pi$ and/or $\pi\pi$ states. In general, one can derive non-perturbative relations between finite-volume matrix elements of local external currents ($\Jc$) and these classes of infinite-volume amplitudes where the electromagnetic contribution is treated perturbatively. In this direction, it has been shown how $\Jc\to 2$, $1+\Jc\to 2$~\cite{Lellouch:2000pv,Briceno:2014uqa,Briceno:2015csa, Briceno:2021xlc, Meyer:2011um,  Feng:2014gba} and $2+\Jc\to 2$~\cite{Briceno:2015tza,Baroni:2018iau}~\footnote{Here, $1$ and $2$ refer to the number of stable hadrons present in the asymptotic states.} transition amplitudes may be constrained from finite-volume matrix elements. Albeit challenging, successful numerical results have already been obtained for various processes~\cite{Briceno:2015dca,Briceno:2016kkp,Alexandrou:2018jbt,Andersen:2018mau,Niehus:2021iin}. In particular, the specific amplitudes that will be needed to be constrained are $\pi \gamma^\star \to \pi$,  $\gamma^\star \to\pi\pi$, and $\pi \pi\gamma^\star \to \pi\pi$.~\footnote{Although to date no calculations of $\pi \pi \gamma^\star \to \pi\pi$ have been obtained via lattice QCD, various non-trivial tests have been presented in the existing formalism~\cite{Briceno:2019nns,Briceno:2020xxs}.}

   The amplitude considered here, which falls under a call of amplitudes we generically label as $\Jc + \Jc \to 2$ processes, involves two currents that are displaced in time, and as a result in general the product can not be considered to be local. Matrix elements of non-local currents are already being studied via lattice QCD. Notable examples include two-photon radiative decays of single-hadron states~\cite{Dudek:2006ut,Gerardin:2016cqj}. The first theoretical attempt to formally understand how long-range processes where intermediate multi-particle states can go on-shell was presented in Ref.~\cite{Christ:2015pwa} in the context of $K^0-\bar{K}^0$ mixing, where there can be intermediate $\pi\pi$ states. This work was later generalized for arbitrary reactions of the form $1+\Jc \to 1+\Jc$~\cite{Briceno:2019opb} where intermediate two particles can go on-shell up to the first three-particle threshold.~\footnote{We point the reader to Ref.~\cite{Briceno:2017max} for a recent review on these topics.} 
   
   The outstanding challenge that was not addressed in Ref.~\cite{Briceno:2019opb} for studying processes of the form of $\Jc + \Jc \to 2$ is associated with the fact that one can not only have on-shell two-particle states between the two currents but also in the final states. This leads to new classes of power-law finite-volume effects that must be isolated. As will be shown in detail, these can be isolated and they depend on physical subprocesses that can be constrained for simpler finite-volume Euclidean correlation functions. In short, we provide an exact formalism that relates the \ggpipi amplitude to a combination of quantities that can be obtained from finite-volume Euclidean correlation functions.

The rest of this manuscript is organized as follows. In \cref{sec:Tdef}, we give an overview of the work and present our main result in Eq.~\eqref{eq:Main_eq}. In \cref{sec:long_range_amplitude}, we review the form of the infinite-volume Minkowski amplitude of interest as derived in Refs.~\cite{Sherman:2022tco}. In~\cref{sec:Main}, we derive the main result mentioned before. In particular, we give the exact linear combination of quantities that can be used to constrain the non-trivial piece of the \ggpipi amplitude. These include the $\pi \pi \to \pi\pi$, $\pi \gamma^\star \to \pi$, and $\gamma^\star \to\pi\pi$ amplitudes and the finite, Euclidean spacetime analogue of the desired Minkowski matrix elements, all of which can be obtained via lattice QCD. In~\cref{sec:FVCorr} we derive the finite volume correction to the amplitude. Finally, in~\cref{sec:conclusions} we provide a summary of this work and an outlook.

%%%%%%%%%%%%%%%%%%%%%%%%%%%%%%%%%%%%%%%%%%%%%%%%%%%%%%%%%%%%%%%%%%%%%%%%
%	Section :: Overview
%%%%%%%%%%%%%%%%%%%%%%%%%%%%%%%%%%%%%%%%%%%%%%%%%%%%%%%%%%%%%%%%%%%%%%%%
\section{Overview}
\label{sec:Tdef}

Here we present an extension of the framework in Ref.~\cite{Briceno:2019opb} in order to access $\Jc + \Jc \to 2$ long-range process from finite-volume matrix elements. We focus on the specific reaction \ggpipi where the $\pi\pi$ final state is projected to some definite angular momentum quantum numbers $J,m_J$. Therefore, we consider only conserved electromagnetic currents $\Jc^{\mu}$. The goal of this work is to derive the formalism that will allow for the determination of this amplitude from lattice QCD correlation functions, for kinematics where two-particle intermediate states may go on-shell.

\begin{figure}[t]
	\begin{center}
	\includegraphics[width=.9\textwidth]{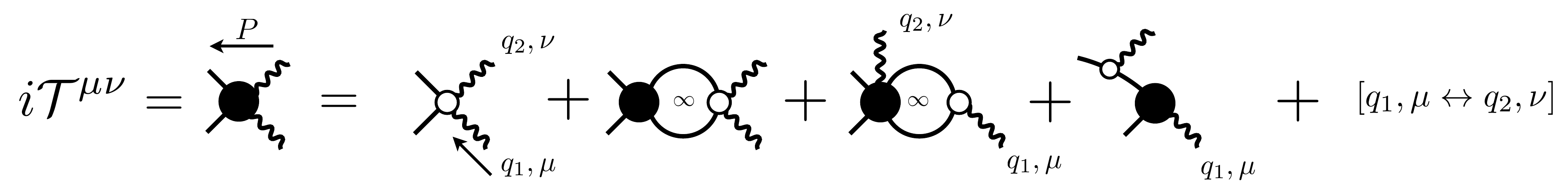}
	\caption{Shown is the diagrammatic definition of the \ggpipi amplitude, defined in Eq.~\eqref{eq:Tggpipi}. The ``wiggly''-lines represent virtual photons with Lorentz indices $\mu$ and $\nu$. Solid lines represent $\pi$ lines. The open circles are kernels, with properties described in the text, and black circles are amplitudes defined subsequently in Fig.~\ref{fig:iM_iH_iW}. The ``$\infty$'' symbol emphasizes that these loops are being evaluated in an infinite volume. The ``$\left[q_1,\mu\leftrightarrow q_2,\nu\right] $'' symbol denotes the presence of diagrams identical to the ones shown except with the labels of the virtual photons swapped.}
	\label{fig:iT}
	\end{center}
\end{figure}

We define the amplitude of interest $\Tc^{\mu\nu}$, depicted diagrammatically in Fig.~\ref{fig:iT}, as the Fourier transform of the corresponding Minkowski-signature time-ordered matrix element of two local currents
\begin{align}
	\label{eq:Tggpipi}
	\Tc^{\mu\nu}(P;q_1) \equiv i\int \diff^4 x \, e^{-i q_1 \cdot x - \epsilon |x^0|} \, \bra{\pi\pi(J,m_J),P;\rm{out}} \mathrm{T} \{ \Jc^{\mu} (x) \Jc^{\nu} (0) \} \ket{\Omega} \, ,
\end{align}
where $P=(E,\textbf{P})$ is the four-momentum of the asymptotic outgoing $\pi\pi$ scattering state, $q_1=(\omega, \textbf{q}_1)$ is the momentum being carried by one of the currents, and $\mathrm{T}$ is the standard time-ordering operator in Minkowski space. Four-momentum conservation requires that the unspecified current carries a momentum $q_2 = P - q_1$. The local conserved currents are functions of the Minkowski spacetime point $x = (x^0,\mathbf{x})$, with one current located at the origin to eliminate the overall momentum-conserving delta function. In Eq.~\eqref{eq:Tggpipi}, we have explicitly introduced $\epsilon$, which should ultimately be taken to zero after integration.~\footnote{Note that we keep the dependence on angular momentum for $\Tc^{\mu\nu}(P;q_1)$ implicit.}

%%%%%%%%%%%%%%%%%%
%	figure
%%%%%%%%%%%%%%%%%%
%%%%%%%%%%%%%%%%%%
%	figure
%%%%%%%%%%%%%%%%%%

As was derived in Ref.~\cite{Sherman:2022tco}, this amplitude can be written in terms of purely on-shell quantities and physical singularities in the form
\begin{align}
	\label{eq:Tmunu}
	i\Tc^{\mu\nu}(P,\hat{\mathbf{p}}^{\star};q_1) \equiv 
	\sum \left\{iw^{\mu}_{\mathrm{on}}\,iD\,i\overline{\Hc}^{\nu}\right\}
    + i\Tc^{\mu\nu}_{\df} (P,\hat{\mathbf{p}}^{\star} ;q_1) \, .
\end{align}
Here we have not yet projected the final state to a definite $J^P$, thus the amplitude depends on the relative orientation of the pions $\hat{\mathbf{p}}^{\star}$ where the $\star$ indicates we evaluate the momentum in the center-of-momentum (CM) frame of the pions. The first term in~\cref{eq:Tmunu} contains simple poles that depend on the $\pi \gamma^\star \to \pi$ and $\gamma^\star \to\pi\pi$ amplitudes (denoted by $w^{\mu}_{\mathrm{on}}$ and $\overline{\Hc}^{\nu}$, respectively), both of which can be accessed via lattice QCD through previously derived formalisms~\cite{Meyer:2011um, Briceno:2015csa, Andersen:2018mau, Feng:2014gba}. A discussion on the specifics of these functions is given in Section~\ref{sec:long_range_amplitude}. In the context of the \ggpipi amplitude, this term represents the pion-pole contribution~\cite{Colangelo:2014pva,Colangelo:2014dfa,Colangelo:2015ama,Danilkin:2019opj,Danilkin:2019mhd,Aoyama:2020ynm}, which we describe in more detail in~\cref{app:decomp}. From the viewpoint of this work, these are previously determined functions, and thus the only quantity that is otherwise unconstrained is $\Tc^{\mu\nu}_{\df}$. The subscript ``$\df$'' stands for \emph{divergence free}, which refers to the fact that the amplitude does not contain any kinematic or spurious singularities, which are encapsulated by the first term of~\cref{eq:Tmunu}, but does contain dynamical singularities such as two-particle production branch cuts, and possible bound and resonance state pole singularities. This subscript appears in other amplitudes introduced below and has a similar definition. 

We then project the decomposition to a definite partial wave in order to link Eq.~\eqref{eq:Tmunu} to the definition \eqref{eq:Tggpipi}. The partial-wave projection can be found by integrating over the solid angle subtended by the $\pi\pi$ CM relative momentum, weighted by an appropriate spherical harmonic,
\begin{align}
	\label{eq:PW}
	\Tc^{\mu\nu}(P;q_1) = \frac{1}{\sqrt{4\pi}} \, \int \! \diff\hat{\mathbf{p}}^{\star} \,  Y_{J m_J}^{*} (\hat{\mathbf{p}}^{\star}) \, \Tc^{\mu\nu}(P,\hat{\mathbf{p}}^{\star};q_1) \, ,
\end{align}
where $m_J$ is the total angular momentum projection along some fixed $z$ axis. The projection of the first term of Eq.~\eqref{eq:Tmunu} will contribute known kinematic singularities similar to those discussed in Ref.~\cite{Briceno:2020vgp}, and the second term simply produces the projection $\Tc^{\mu\nu}_{\df}(P;q_1)$. In Section~\ref{sec:long_range_amplitude} we outline an exact definition of this quantity in terms  of scattering amplitudes of physical subprocesses, known kinematic functions, and unknown non-singular dynamical functions that are purely real.

The central result of this work relates the infinite volume amplitude $\Tc_{\df}^{\mu\nu}$ to finite-volume Euclidean-signature matrix elements which we can compute, e.g., using lattice QCD. We work in a finite-cubic-volume with side length $L$ and periodic boundary conditions, which imposes that spatial momenta are quantized as $\mathbf{P} = 2\pi \mathbf{n} / L$ where $\mathbf{n} \in \mathbb{Z}^3$. We assume that the size of the volume is such that $mL \gg 1$, where $m$ is the mass of the pion. We define the finite-volume Euclidean spacetime matrix element which closely resembles the infinite-volume counterpart of \cref{eq:Tggpipi} as
\begin{align}
	\label{eq:GL}
	M_L^{\mu\nu}(\tau,P_n; \mathbf{q}_1)
	\equiv
	\int_L\! \diff^3 \mathbf{x}  \,  e^{i \mathbf{q}_1 \cdot  \mathbf{x} }  \, \bra{P_n , L} \mathrm{T}_E \{ \Jc^{\mu}_E (\tau, \mathbf{x}) \Jc_E^{\nu} (0) \} \ket{\Omega} \, ,
\end{align}
where $\tau$ is Euclidean time. The matrix element depends on the finite volume spectrum $E_n$ for the $n$th state at a given total momentum $\mathbf{P}$, with $P_n = (E_n,\mathbf{P})$.~\footnote{Note that the spectrum implicitly depends on the volume and momentum, $E_n = E_n (L,\mathbf{P})$, as well as other quantum numbers such as those dictated by the representation of the system under the cubic group.} Finite-volume eigenstates are normalized to unity, 
\begin{align}
	\braket{P_m,L | P_n',L} = \delta_{mn} \delta_{\mathbf{P}\mathbf{P}'} \, , \nn 
\end{align}
which differs from the relativistic normalization of infinite-volume single particle states.~\footnote{Infinite-volume single particle states are normalized as $\braket{\mathbf{p}'|\mathbf{p}} = (2\pi)^3 \, 2\omega_{\mathbf{p}} \, \delta^{(3)}(\mathbf{p}' - \mathbf{p})$ where $\omega_{\mathbf{p}} = \sqrt{m^2 + \mathbf{p}^2}$.} The local currents $\Jc_E$ here translate in Euclidean time as 
\begin{align}
	\Jc_{E}^{\mu}(\tau,\mathbf{x}) = e^{H_L\tau} \, \Jc^{\mu}(0,\mathbf{x}) \, e^{-H_L\tau} \, , \nn 
\end{align}
where $H_L$ is the finite-volume QCD Hamiltonian, and $\mathrm{T}_E$ is the Euclidean time-ordering operator, which orders the current insertion depending on the values of $\tau$ and $0$.

Although computationally challenging, these matrix elements can in principle be computed from time-displaced Euclidean-signature correlation functions. This is the case in lattice QCD studies of two-photon radiative decays of QCD-stable hadrons~\cite{Dudek:2006ut,Gerardin:2016cqj}. Since we are ultimately interested in momentum-space long-range matrix elements, which do not depend on the temporal signature, we should integrate Eq.~\eqref{eq:GL} over Euclidean time $\tau$ with an exponential weight that depends on the desired frequency of the external current, i.e. $\int \diff\tau \, e^{\omega \tau}$. Unfortunately, the resulting integral is in general not equivalent to the Fourier transform of the Minkowski spacetime matrix element $\Tc^{\mu\nu}(P;q_1)$ for two reasons. First, the finite-volume states are not exponentially close to the infinite-volume ones. Second, depending on the kinematics of the external states and the current, the resulting integral does not converge. We first address the second issue, which was discussed in Ref.~\cite{Briceno:2019opb}. The first of these issues amounts to finite-volume corrections which are discussed in Section~\ref{sec:FVCorr}, while the second is the subject of Section~\ref{sec:Main}.

Intuitively, the integral fails to converge when intermediate states created in the long-range process can go on-shell. Fortunately, the contribution from these intermediate states can be determined completely from the spectrum and matrix elements of the physical subprocesses. In particular, in the long-range process of interest, \ggpipi, the physical subprocess that can lead to intermediate on-shell states is  $\gamma^\star\to \pi\pi \to \gamma^\star \pi\pi$. In other words, one must determine the $\pi\pi$ spectrum, extract the $\pi\pi\to\pi\pi$ scattering amplitude~\cite{Luscher:1991n1, Rummukainen:1995vs, Kim:2005gf}, and the finite-volume matrix elements corresponding to $\gamma^\star\to \pi\pi $~\cite{Meyer:2011um, Briceno:2015csa} and $\pi\pi \to \gamma^\star \pi\pi$~\cite{Briceno:2015tza, Baroni:2018iau} processes. These three subprocesses are diagrammatically defined in Fig.~\ref{fig:iM_iH_iW}. Reference~\cite{Briceno:2019opb} proposed that these divergent contributions be removed from the $\tau$-dependent correlation function. We label the divergent contribution as $M_L^{\mu\nu,<}$, where ``$<$'' indicates that the contribution from a number of states that lie below some finite cutoff is included in the subtraction.

Following Ref.~\cite{Briceno:2019opb}, we define the subtracted Euclidean time-dependent matrix element, $M_L^{\mu\nu,>}$, as
\begin{align}
	\label{eq:G>}
	M_L^{\mu\nu,>}(\tau,P_n; \textbf{q}_1) \equiv M_L^{\mu\nu}(\tau,P_n; \mathbf{q}_1) - M_L^{\mu\nu,<}(\tau,P_n; \mathbf{q}_1).
\end{align}
This function does not contain any on-shell processes, and thus its integral over $\tau$ converges. The resulting correlation function, although unphysical, is insensitive to the time signature used to access it. This means, that it coincides with a contribution to the desired Minkowski correlation function, and it can be qualitatively understood as \emph{the short-distance} contribution to the quantity $\Tc_{L}^{\mu\nu}$, the finite-volume analogue of Eq.~\eqref{eq:Tggpipi}. After this procedure, we restore the long-distance modes from the divergent terms which depend on the aforementioned $\gamma^\star\to \pi\pi $ and $\pi\pi \to \gamma^\star \pi\pi$ matrix elements and the $\pi\pi$ spectrum. As detailed in Section~\ref{sec:Main}, this can be made explicit by summing over simple poles associated with finite-volume states, and we denote this contribution as $\Tc_L^{\mu\nu, <}$. Because this last contribution is in frequency-space, the resulting correlation function can be directly linked to the Minkowski quantity of interest, Eq.~\eqref{eq:Tggpipi}.

Since the $\Tc_L^{\mu\nu, <}$ term depends explicitly on the finite-volume spectrum, it has power-law finite-volume artifacts which we need to remove in order to identify the relationship to the desired $\Tc^{\mu\nu}_{\df}$. The procedure is similar to that outlined in Ref.~\cite{Briceno:2019opb}, where we sum the finite-volume correlation function to all orders, and isolate contributions that lead to power-law volume-dependent effects. Since we work with two-particle final states, as opposed to the single-particle states of Ref.~\cite{Briceno:2019opb}, additional corrections arise from their rescattering. In Section~\ref{sec:FVCorr}, a complete derivation of the isolation of the finite-volume correction to $\Tc^{\mu\nu}_{\df}$ is given. Combining this result with the aforementioned separation of long-distance modes in the Euclidean matrix element, our result can be succinctly summarized by the following relation,
\begin{align}
	\label{eq:Main_eq}
	\frac{\rrvec}{\sqrt{2 E_n L^3}}
	\cdot
	\Tc^{\mu\nu}_{\df}(P_n;q_1)
	& = 
	\int \diff \tau  \, e^{\omega\tau}
	\, M_L^{\mu\nu,>}(\tau,P_n; \textbf{q}_1) + \left[ \, 
	\Tc_L^{\mu\nu, <}(P_n;q_1)
	+
	\frac{\rrvec}{\sqrt{2 E_n L^3}} \cdot 
	\Delta \Tc^{\mu\nu}_{L,\df}(P_n;q_1) \, \right] \, ,
\end{align}
which holds at the finite-volume energies $E_n$ and ignores corrections which scale like $\mathcal{O}(e^{-mL})$. Here we introduce $\rrvec$, which is related to the well-known Lellouch-L\"uscher factor~\cite{Lellouch:2000} as first written in Ref.~\cite{Briceno:2014uqa}. These factors serve to correct finite-volume effects arising from the final $\pi\pi$ state, and we provide an exact definition in~\cref{eq:Rvec} where we follow the notation introduced in Ref.~\cite{Briceno:2021xlc}. The additive correction, $\Delta \Tc^{\mu\nu}_{L,\df}$, is the desired finite-volume correction which depends on the physical quantities associated with the $\pi\gamma^\star\to \pi $, $\gamma^\star\to \pi\pi $, and $\pi\pi\gamma^\star  \to \pi\pi$ subprocesses. An exact expression of $\Delta \Tc^{\mu\nu}_{L,\df}$ is given in Eq.~\eqref{eq:deltaTLdf}. We group the terms $\Tc_L^{\mu\nu,<}$ and $\Delta \Tc_{L,\df}^{\mu\nu}$ to emphasize the fact that these terms must use the same parametrization used for $\Delta \Tc$ because otherwise these poles will not exactly cancel, and can lead to unphysical behavior of the amplitude.

%%%%%%%%%%%%%%%%%%%%%%%%%%%%%%%%%%%%%%%%%%%%%%%%%%%%%%%%%%%%%%%%%%%%%%%%
%	Section :: On-shell representation of amplitude
%%%%%%%%%%%%%%%%%%%%%%%%%%%%%%%%%%%%%%%%%%%%%%%%%%%%%%%%%%%%%%%%%%%%%%%%
\section{On-shell representation of the amplitude}
\label{sec:long_range_amplitude}

\begin{figure}[t]
	\begin{center}
	\includegraphics[width=1\textwidth]{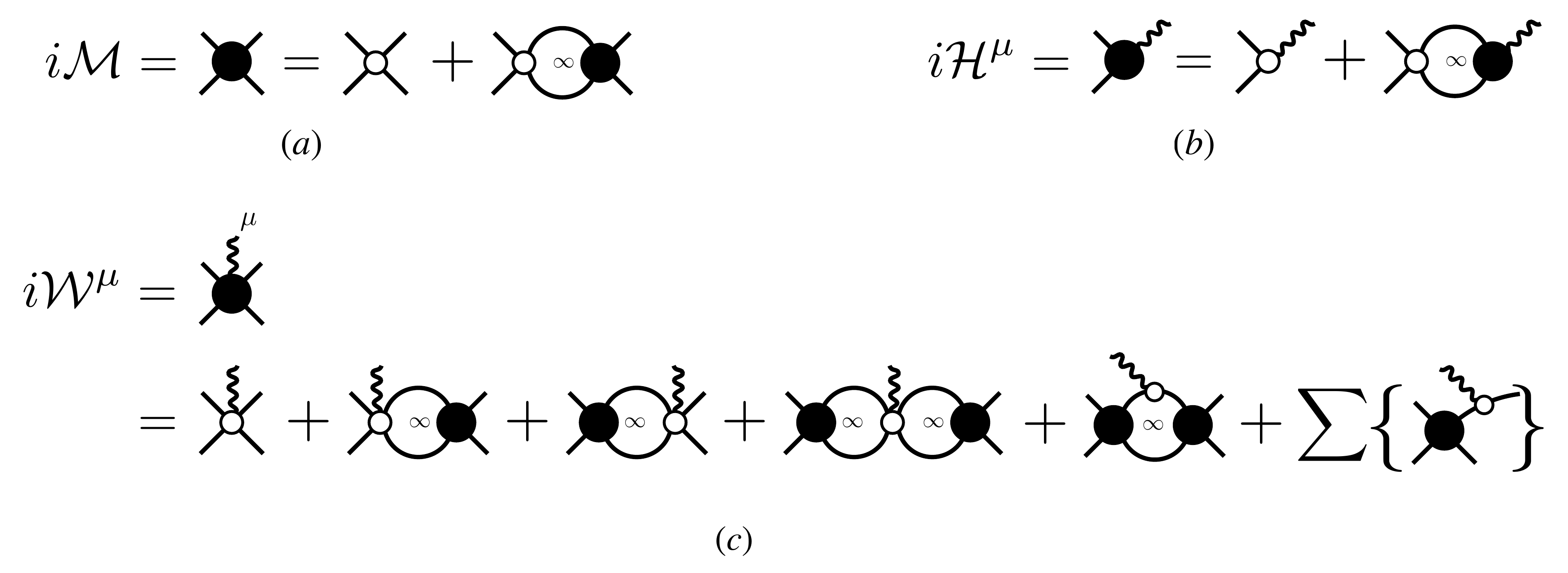}
	\caption{Shown are the diagrammatic representation of the (a) $\pi\pi\to\pi\pi$, (b) $\gamma^\star\to\pi\pi$, and (c) $\pi\pi\gamma^\star\to\pi\pi$ amplitudes and their all-orders expressions. The open circles represent real-valued, non-singular functions, whose properties are discussed in the main text. As discussed in detail in the text, the ``$\sum$'' symbol represents the sum over all possible insertion of the current in the external legs. An example of this is shown explicitly for the pole piece of $\Tc$ is shown diagrammatically in Fig.~\ref{fig:T_poles}.}
	\label{fig:iM_iH_iW}
	\end{center}
\end{figure}

We briefly review the general on-shell formalism presented in Ref.~\cite{Sherman:2022tco} for the specific amplitude of interest, $\Tc^{\mu\nu}$. Generally, the amplitude can be written to all orders within some generic relativistic effective field theory as a self-consistent equation in terms of short-distance kernels which do not have singularities for the kinematic range of interest. This self-consistent equation is shown diagrammatically in Fig.~\ref{fig:iT} for $i\Tc^{\mu\nu}$, where the kernels are shown with white open circles connected to electromagnetic currents denoted by ``wiggly'' lines. The kernels are integrated over four-dimensional momentum loops of pion propagators (shown by solid straight lines) with $2 \to 2$ scattering and $2 + \Jc \to 2$ transition amplitudes, called $\Mc$ and $\Wc^\mu$, respectively. These amplitudes are subsequently defined in terms of their own self-consistent equations in Fig.~\ref{fig:iM_iH_iW}. The amplitude $\Tc^{\mu\nu}$ also contains a contribution from the $\Jc \to 2$ production amplitude, called $\Hc^\mu$, where one of the pions subsequently interacts with the other current. This interaction leads to a new kernel associated with the one-body matrix element $1 + \Jc \to 1$ transition. Note that in Fig.~\ref{fig:iT} we have only shown the diagrams where the currents are inserted in a particular time order and have left the other time ordering implicit, as indicated by ``$\left[q_1,\mu\leftrightarrow q_2,\nu\right] $''.

Reference~\cite{Sherman:2022tco} shows the projection of these generic relations to their on-shell form, which separates all long-range intermediate state singularities from all short-distance physics which are absorbed into unknown non-singular functions. These results hold for a kinematic region where only two-particle states can be produced in the kinematic variable $P^2$, and the virtualities of the currents are such that higher-number particle production thresholds are prohibited. This procedure results in Eq.~\eqref{eq:Tmunu}, and the remainder of this section details each of the building blocks.

%%%%%%%%%%%%%%%%%%
%	figure
%%%%%%%%%%%%%%%%%%
%%%%%%%%%%%%%%%%%%
%	figure
%%%%%%%%%%%%%%%%%%

The first term, $iw^{\mu}_{\mathrm{on}}\, iD\, i\overline{\Hc}^{\nu}$, includes infrared pole singularities associated with individual currents coupling to one of the external particles. Here, $iw^{\mu}_{\mathrm{on}}$ refers to a specific definition of the single particle matrix elements for $\pi \gamma^{\star} \to \pi$ introduced in Ref.~\cite{Baroni:2018iau}. In particular, this definition places the single-particle form factors on-shell, while allowing for the kinematic prefactors to be off-shell. For charged pions, this is explicitly given by
\begin{align}
	\label{eq:won}
	iw^{\mu}_{\mathrm{on}}(k_f ; k_i) = (k_f + k_i)^\mu \, if\left( Q^2 \right) \, ,
\end{align}
where $f $ is the physical form factor, $Q^2=-(k_f-k_i)^2$ is the momentum-transfer-squared to the pion, and $k_i$ ($k_f$) denote the incoming (outgoing) momentum of the pion, respectively. In this framework, $k_i^2$ and $k_f^2$ need not coincide with $m^2$. The next quantity that appears in the first term of Eq.~\eqref{eq:Tmunu} is $iD$, which is the pole piece of the single pion propagator defined as
\begin{align}
	\label{eq:D_bareprop}
	iD(k) \equiv \frac{i}{k^2-m^2+i\epsilon} \, .
\end{align}

\begin{figure}[t]
	\begin{center}
	\includegraphics[width=0.8\textwidth]{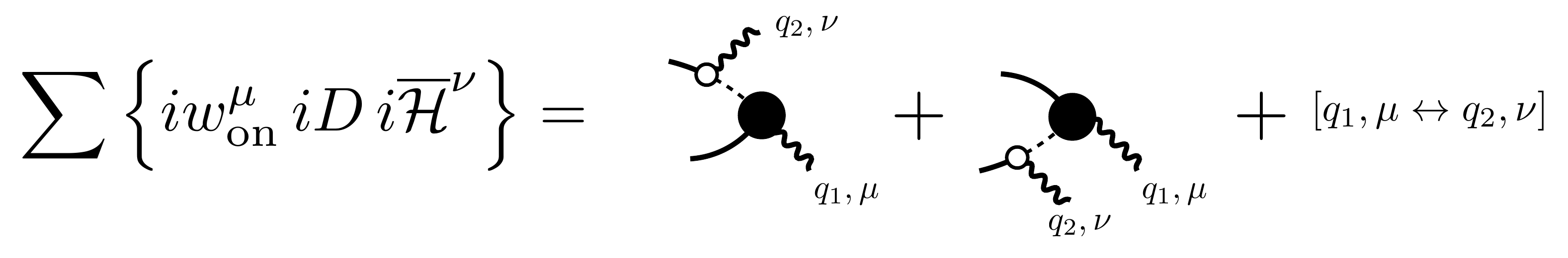}
	\caption{The kinematically divergent term, Eq.~\eqref{eq:T_poles}, for the $\gamma^{\star}\gamma^{\star} \to \pi^+\pi^-$ amplitude of Eq.~\eqref{eq:Tmunu}. The dashed line indicates the pole piece of the propagator $iD$ in Eq.~\eqref{eq:D_bareprop}.}
	\label{fig:T_poles}
	\end{center}
\end{figure}

The last piece in the divergent term of Eq.~\eqref{eq:Tmunu} is $\overline{\Hc}^{\mu}$, which is closely related to the $\gamma^\star\to \pi\pi$ transition amplitude ${\Hc}^{\mu}$.~\footnote{Although the process $\pi \gamma^{\star} \to \pi$ is related to the crossed channel $\gamma^{\star} \to \pi\pi$, we use different symbols to indicate that in our kinematic region of interest, these amplitudes produce different physical effects.} We depict the self-consistent relation for ${\Hc}^{\mu}$ in Fig.~\ref{fig:iM_iH_iW}(b). The difference between $\overline{\Hc}^{\mu}$ and ${\Hc}^{\mu}$ is the addition of kinematic barrier factors in the definition of $\overline{\Hc}^{\mu}$. These barrier factors cancel in the limit where the particle coupling to the current goes on-shell. In Appendix~\ref{app:decomp}, we give an explicit definition of $\overline{\Hc}^{\mu}$. Note that associated with the divergent piece is the summation notation ``$\sum$''. As discussed in Ref.~\cite{Briceno:2020vgp, Sherman:2022tco}, this indicates a sum over all external legs where the current could be inserted. For example, if the final state is composed of $\pi^+\pi^-$, the electromagnetic current can couple to both the $\pi^+$ and the $\pi^-$, resulting in four such contributions when accounting for the [$q_1,\mu \leftrightarrow q_2,\nu$] crossed channel. For this scenario, the kinematically divergent term is depicted in Fig.~\ref{fig:T_poles} and is explicitly given by
\begin{align}
	\label{eq:T_poles}
	\sum \left\{iw^{\mu}_{\mathrm{on}}\,iD\,i\overline{\Hc}^{\nu}\right\} & = iw_{\mathrm{on},+}^{\nu}(P-p;q_1-p) \, iD(q_1-p) \, i\overline{\Hc}^{\mu}(\mathbf{q}_1-\mathbf{p};q_1) \, \nn  \\[5pt]
	& \qquad + iw_{\mathrm{on},-}^{\nu}(p;p-q_2) \, iD(p-q_2) \, i\overline{\Hc}^{\mu}(\mathbf{p}-\mathbf{q}_2;q_1) + [q_1,\mu \leftrightarrow q_2,\nu] \, ,
\end{align}
where $w_{\mathrm{on},+}$ and $w_{\mathrm{on},-}$ are the matrix elements for the $\pi^+$ and $\pi^-$, respectively, which are on their mass-shell $(P-p)^2 = p^2 = m^2$. In Appendix~\ref{app:decomp}, we explain in detail how this term can be projected to definite angular momentum.

Having defined the kinematically divergent term in Eq.~\eqref{eq:Tmunu}, we now move toward the description of the divergence-free amplitude $\Tc^{\mu\nu}_{\df}$, which is the object we need to constrain from the finite-volume formalism. It was derived in Ref.~\cite{Sherman:2022tco} that $\Tc^{\mu\nu}_{\df}$ has an on-shell representation given by
\begin{align}
	\label{eq:Tmunu_df}
	i\Tc^{\mu\nu}_{\df} (P;q_1)
	\equiv 
    i\Wc^\mu_{\df}(P;q_2) \, \Ac^{\nu}_{20}(q_2) 
    +
    i\Wc^\nu_{\df}(P;q_1) \, \Ac^\mu_{20}(q_1) 
    +
    i\Mc(P^2) \, \Bc^{\mu\nu}_{20}(P;q_1) \, ,
\end{align}
where $q_2 = P - q_1$, $\Mc$ is the purely hadronic $\pi\pi \to \pi\pi$ partial wave scattering amplitude, and $\Wc_\df$ is related to the $\pi\pi\gamma^\star\to \pi\pi$ amplitude, denoted as $\Wc^\mu$. We show diagrammatic depictions of these amplitudes, as well as their self-consistent equations, in Figs.~\ref{fig:iM_iH_iW} (a) and (c) for $\Mc$ and $\Wc^\mu$, respectively. In particular, Ref.~\cite{Briceno:2020vgp} showed that the full $\pi\pi\gamma^\star\to \pi\pi$ amplitude $\Wc^\mu$ satisfies an equation similar to Eq.~\eqref{eq:Tmunu}, 
\begin{align}
	\label{eq:Mmu}
	i\Wc^{\mu}(P_f,\hat{\mathbf{p}}^{\star}_f;P_i,\hat{\mathbf{p}}^{\star}_i) 
	\equiv 
	\sum \left\{ iw^{\mu}_{\mathrm{on}}\,iD\, i\overline{\Mc} \right\}
	+
	i\Wc^{\mu}_{\df} (P_f,\hat{\mathbf{p}}^{\star}_f;P_i,\hat{\mathbf{p}}^{\star}_i) \, ,
\end{align}
where the first term, similar to the one in Eq.~\eqref{eq:Tmunu}, accounts for explicit kinematic singularities from the current probing the external state with $\overline{\Mc}$ related to the hadronic amplitude $\Mc$ up to the same barrier factors needed in the definition of $\overline{\Hc}^\mu$. Here, $P_i$ and $P_f$ are the total initial and final state momentum flows for this process, while $\hat{\mathbf{p}}_i^{\star}$ and $\hat{\mathbf{p}}_f^{\star}$ are the relative CM orientations for the initial and final state, respectively. The partial wave $\Wc_{\df}^{\mu}(P_f;P_i)$ is found similar to that of Eq.~\eqref{eq:PW}, except we need to project both the initial and final states. The appearance of two terms associated with the $\pi\pi\gamma^{\star} \to \pi\pi$ accounts for the $[q_1,\mu \leftrightarrow q_2,\nu]$ crossing in the amplitude.

%%%%%%%%%%%%%%%%%%
%	figure
%%%%%%%%%%%%%%%%%%
%%%%%%%%%%%%%%%%%%
%	figure
%%%%%%%%%%%%%%%%%%

Every term in Eq.~\eqref{eq:Tmunu_df} contains a real-valued, non-singular~\footnote{Up to potential trivial barrier factors for higher partial waves, see Refs.~\cite{Briceno:2020vgp, Sherman:2022tco} for discussions.} short-distance function which we denote as $\Ac^\mu_{20}$ and $\Bc^{\mu\nu}_{20}$. Each $\Ac^{\mu}_{20}$ can be constrained from $\gamma^\star\to \pi\pi$ transitions~\cite{Briceno:2015csa}, and are defined by the on-shell form for $\Hc^\mu$, e.g. for the $q_1$ kinematic variable,
\begin{align}
	\label{eq:H_on}
	\Hc^\mu (q_1) = \Mc\left(q_1^2\right) \, \Ac^\mu_{20}(q_1) \, .
\end{align}
Therefore, we will treat $\Ac^{\mu}_{20}$ as a ``known'' function which is determined from a previously defined formalism as given in Ref.~\cite{Meyer:2011um, Briceno:2015csa}, leaving $\Bc^{\mu\nu}_{20}$ as the only unconstrained function in $\Tc^{\mu \nu}_{\df}$. Therefore, coupled with the finite-volume framework presented in this work, this unknown dynamical function can be constrained from lattice QCD studies. 

The $\Mc$ and $\Wc^\mu_{\df}$ amplitudes can similarly be written in terms of known kinematic singular functions and unknown real, non-singular functions, see Ref.~\cite{Briceno:2020vgp} for a detailed discussion. For $\Mc$, it is well known that the unknown dynamical contribution can be written in terms of the K matrix, or equivalently phase shifts and mixing angles, while the kinematic singularities are encoded in the phase-space of the two-particle system. Reference~\cite{Briceno:2020vgp}, derived the on-shell representation for the $\Wc^{\mu}_{\df}$ amplitudes, where it was found that in addition to phase-space singularities, $\Wc^{\mu}_{\df}$ also contains triangle singularities which are isolated exactly. Again, from the perspective of this work, the $\Mc$ and $\Wc^\mu$ amplitudes are ``known'', and therefore the only new object that our framework constrains in $\Tc^{\mu\nu}_{\df}$ is $\Bc^{\mu\nu}_{20}$. Once this object is determined, then combining it with the various on-shell relations in this section provides a complete description of the $\Jc + \Jc \to 2$ amplitude.

While we have simplified the problem to one of determining a real-valued non-singular function $\Bc_{20}^{\mu\nu}$, in general, it is still a complicated object due to the Lorentz structure. It is useful to decompose the object, and even the full amplitude $\Tc^{\mu\nu}$, into scalar form-factors associated with some known kinematic Lorentz structure, the form of which depends on the final state quantum numbers. Therefore, we can further reduce the problem to that of finding the individual scalar form-factors. Appendix~\ref{app:decomp} provides a summary of the Lorentz decomposition for the amplitude, as well as an explicit example for the $J^P = 0^+$ sector.

As a final comment, the $\Tc^{\mu\nu}$ amplitude must satisfy the Ward-Takahashi identity for conserved vector currents. It was shown in Ref.~\cite{Sherman:2022tco} that this identity places a constraint on the unknown $\Bc^{\mu\nu}_{20}$ function in terms of the other lower-point functions in the limit where one of the photon momenta vanishes. We do not state the constraint here, and point the reader to Ref.~\cite{Sherman:2022tco} for details, but note that this constraint can aid in developing parameterizations for the $\Bc^{\mu\nu}_{20}$ object for numerical applications as they must all collapse to the same result in the vanishing photon momentum limit.

%%%%%%%%%%%%%%%%%%%%%%%%%%%%%%%%%%%%%%%%%%%%%%%%%%%%%%%%%%%%%%%%%%%%%%%%
%	Section :: Relating Euclidean and Minkowski matrix elements
%%%%%%%%%%%%%%%%%%%%%%%%%%%%%%%%%%%%%%%%%%%%%%%%%%%%%%%%%%%%%%%%%%%%%%%%
\section{Relating Euclidean and Minkowski matrix elements}
\label{sec:Main}

In this section, we summarize the details of the procedure which relates the Euclidean matrix element, Eq.~\eqref{eq:GL}, to $\Tc^{\mu\nu}_{L,\df}$. For convenience, we repeat the definition of Eq.~\eqref{eq:GL} here
\begin{align}
	M_L^{\mu\nu}(\tau,P; \mathbf{q}_1)
	\equiv
	\int_L\! \diff^3 \mathbf{x}  \,  e^{i \mathbf{q}_1 \cdot  \mathbf{x} }  \, \bra{P , L} \mathrm{T}_E \{ \Jc^{\mu}_E (\tau, \mathbf{x}) \Jc_E^{\nu} (0) \} \ket{\Omega} \, , \nn 
\end{align}
where we recall that the finite-volume states $\ket{P,L}$ have the quantum numbers of two-pions and have unit normalization. For notational convenience, we write the momentum of the state $P = (E,\mathbf{P})$ instead of $P_n = (E_n, \mathbf{P})$ as we had in Section~\ref{sec:Tdef}, but we stress that it has the same interpretation as a quantized spectrum. We noted that a trivial frequency-dependent weighted integration over the Euclidean time $\tau$ does not converge due to long-range modes in the kinematic region of on-shell processes of interest. Therefore, a trivial identification of the Fourier transformed matrix element $\Tc^{\mu\nu}$ is not possible. Following Ref.~\cite{Briceno:2019opb}, the resolution involved removing the divergent contributions to render the integral convergent and reintroducing these modes via a spectral reconstruction.

The divergent contribution is denoted as $M_L^{\mu\nu,<}$, where we remind the reader that the symbol ``$<$'' indicates that only a ``small'' subset of poles have been calculated explicitly. Particularly, $M_L^{\mu\nu,<}$ can be written using the spectral decomposition as
\begin{align}
	\label{eq:GLtau}
	M_L^{\mu\nu,<}(\tau,P; \mathbf{q}_1)
	\equiv 
	\sum_{n=0}^{N-1} c_n^{\mu\nu} \, \Theta(\tau) \, e^{-[E_n(L,\mathbf{P}-\mathbf{q}_1)-E]|\tau|}
	+
	\sum_{\overline{n}=0}^{\overline{N}-1} \overline{c}_{\overline{n}}^{\nu\mu} \, \Theta(-\tau) \, e^{-E_{\overline{n}}(L,\mathbf{q}_1) |\tau|},
\end{align}
where $\ket{n,L}$ and $\ket{\overline{n},L}$ denote arbitrary finite-volume states satisfying $\braket{\bar n,L|n,L}=\delta_{\bar n n}$, $\Theta$ is the Heaviside step function, $n$ and $\bar{n}$ are integers enumerating the possible discrete states, $E_n$ and $E_{\overline{n}}$ are their respective energies, and $c_n^{\mu\nu}$ and $\overline{c}_{\overline{n}}^{\nu\mu}$ are their corresponding matrix elements, 
\begin{align}
	c_n^{\mu\nu} & \equiv
	\int\! \diff^3 \mathbf{x} \, e^{i \mathbf{q}_1 \cdot  \mathbf{x} } \, 
	\bra{P,L}  \Jc^{\mu} (0,\textbf{x})\ket{n,L}
	\bra{n,L} \Jc^{\nu} (0) \ket{\Omega} \, , \\[5pt]
	\overline{c}_{\overline{n}}^{\nu\mu} & \equiv
	\int\! \diff^3 \mathbf{x} \, e^{i \mathbf{q}_1 \cdot  \mathbf{x} } \, 
	\bra{P,L}  \Jc^{\nu}(0)\ket{\overline{n},L} 
	\bra{\overline{n},L}\Jc^{\mu} (0,\textbf{x}) \ket{\Omega} \, .
\end{align}
Note, that we suppress the total momentum dependence of the $\pi\pi$ finite-volume states for simplicity. The minimal number of terms needed to be included in the sums of Eq.~\eqref{eq:GLtau} is defined by those states that can go-shell in the desired kinematics, which we denote by $N$ and $\overline{N}$. In principle, one can include even more states. But the critical point is that $M_L^{\mu\nu,<}$ can be completely defined in terms of the low-lying spectra and finite-volume spectra that can be determined from two- and three-point functions involving a single current insertion.

With this spectral definition, we define the subtracted time-dependent matrix element $M_L^{\mu\nu,>}$ in Eq.~\eqref{eq:G>}, repeated here for the reader
\begin{align}
	M_L^{\mu\nu,>}(\tau,P; \mathbf{q}_1) \equiv M_L^{\mu\nu}(\tau,P; \mathbf{q}_1) - M_L^{\mu\nu,<}(\tau,P; \mathbf{q}_1) \, . \nn 
\end{align}
We make use of this definition to write the finite-volume Minkowski amplitude, $ \Tc_L^{\mu\nu}$, in terms of the integral of $M_L^{\mu\nu,>}$ and an additive piece that depends on the same low-lying spectra and matrix elements that are necessary for defining $M_L^{\mu\nu,<}$. We derive this identity starting from the definition of $\Tc_L^{\mu\nu}$,
\begin{align}
	\label{eq:TLrel}
	\Tc_L^{\mu\nu}(P;q_1) & =
 	i \int_L \diff^4 x \, e^{-i q_1 \cdot x-\epsilon |x^0|} \, \bra{P,L} \mathrm{T} \{\Jc^{\mu} (x) \Jc^{\nu} (0) \} \ket{\Omega} \, , \nn \\[5pt]
 	& \equiv \Tc_L^{\mu\nu , >}(\omega,P; \textbf{q}_1) + \Tc_L^{\mu\nu , <}(\omega,P; \textbf{q}_1) \, ,
\end{align}
where we have separated the short-and long-distance modes, defining the short-distance contribution $\Tc_{L}^{\mu\nu, >}$ as 
\begin{align}
	\Tc_L^{\mu\nu , >}(\omega,P; \textbf{q}_1) & \equiv 
	i \int_L \diff^4 x \, e^{-i q_1 \cdot x-\epsilon |x^0|} \,
	\bigg[\,
	\bra{P,L}  \mathrm{T} \{\Jc^{\mu} (x) \Jc^{\nu} (0) \} \ket{\Omega} \nn\\[5pt]
	&\hspace{4cm} -
	\sum_{n=0}^{N-1} 
	\bra{P,L}   \Jc^{\mu} (x) \ket{{n},L}
	\bra{{n},L}\Jc^{\nu} (0) \ket{\Omega}\Theta(t) \nn\\[5pt]
	&\hspace{4cm} -
	\sum_{\overline{n}=0}^{\overline{N}-1} 
	\bra{P,L}   \Jc^{\nu} (0) \ket{\overline{n},L}
	\bra{\overline{n},L}\Jc^{\mu} (x) \ket{\Omega}\Theta(-t)
	\, \bigg] \, , \nn \\[5pt]
	& = \int\! \diff \tau \, e^{\omega \tau} \, \left[ \, M_L^{\mu\nu}(\tau,P; \textbf{q}_1) - M_L^{\mu\nu,<}(\tau,P; \textbf{q}_1) \, \right] \, ,
\end{align}
where the second line follows from evaluating the spatial integral analytically, Wick-rotating to Euclidean time, and identifying the matrix elements we have defined in Eq.~\eqref{eq:GL} and~\eqref{eq:GLtau}. With the short-distance modes encapsulated by $\Tc_L^{\mu\nu, >}$, the remaining terms constitute the long-range contribution $\Tc_L^{\mu\nu, <}$ arising from low-lying states
\begin{align}
	\Tc_L^{\mu\nu, <}(P;q_1) & \equiv
	\int_{-\infty}^\infty \!\diff t
	\, e^{-i\omega x^0-|t|\epsilon}
	\left[ \,
	\sum_{n=0}^{N-1}
	c_n^{\mu\nu} \, \Theta(t) \, 
	e^{-i[E_n(L,\mathbf{P}-\mathbf{q}_1)-E]t}
	+
	\sum_{\overline{n}=0}^{\overline{N}-1}
	\overline{c}_{\overline{n}}^{\nu\mu} \, \Theta(-t) \,
	e^{iE_{\overline{n}}(L,\mathbf{q}_1) t}
	\, \right]_{\epsilon =0} \nn\\[5pt]
	& = \sum_{n=0}^{N-1}
	\frac{c_n^{\mu\nu}}{E_n(L,\mathbf{P}-\mathbf{q}_1)-(E-\omega)}
	+
	\sum_{\overline{n}=0}^{\overline{N}-1}
	\frac{\overline{c}_{\overline{n}}^{\nu\mu} }{E_{\overline{n}}(L,\mathbf{q}_1) - \omega} \, .
\end{align}
This separation follows from the spectral representation for the Minkowski matrix element, similar to the Euclidean case shown in Eq.~\eqref{eq:GLtau}, and isolating the high-energy states that can not go on shell. Subsequently, we evaluated the spatial integrals, and in the long-range contribution $\Tc_L^{\mu\nu, <}$ we performed the final temporal integral.

Equation~\eqref{eq:TLrel} relates the Euclidean-signature matrix elements, in terms of computable correlation functions and spectral reconstructions of the low-lying modes. Up to this point, the steps needed are similar to those considered in Ref.~\cite{Briceno:2019opb} for simpler processes. As one would expect, the final step, which amounts to corrections for the power-law finite-volume artifacts is specific to the process of interest, and will necessarily differ from the expression found in Ref.~\cite{Briceno:2019opb}. In particular, $1+\Jc \to 1+\Jc$ amplitudes have single-particle final states, so there is no finite-volume correction coming from these objects. However, in this work we are dealing with a two-particle final state, thus the relation between $\Tc^{\mu\nu}_L$ and $\Tc^{\mu\nu}_{\df}$ must know about the finite-volume effects of these final states. In Section~\ref{sec:FVCorr} we show that $\Tc^{\mu\nu}_L$ satisfies
\begin{align}
	\Tc^{\mu\nu}_L(P_n;q_1) = \frac{\rrvec}{\sqrt{2 E_n L^3}} \cdot
	\bigg( \, \Tc^{\mu\nu}_{\df}(P_n;q_1) -\Delta \Tc^{\mu\nu}_{L,\df}(P_n;q_1) \, \bigg) \, ,
\end{align}
where as mentioned in Section~\ref{sec:Tdef}, $\rrvec$ is related to the well-known Lellouch-L\"uscher factor~\cite{Lellouch:2000} and $\Delta \Tc^{\mu\nu}_{L,\df}$ is the finite-volume correction, both of which are defined in Section~\ref{sec:FVCorr}. We have also reintroduced $P_n = (E_n,\mathbf{P})$ as this relation only holds at the finite-volume spectrum. Combining all the pieces, we arrive at the main result of this work,
\begin{align}
	\frac{\rrvec}{\sqrt{2 E_n L^3}}
	\cdot
	\Tc^{\mu\nu}_{\df}(P_n;q_1)
	& = 
	\int \diff \tau  \, e^{\omega\tau}
	\, M_L^{\mu\nu,>}(\tau,P_n; \textbf{q}_1) +
	\Tc_L^{\mu\nu, <}(P_n;q_1)
	+
	\frac{\rrvec}{\sqrt{2 E_n L^3}} \cdot 
	\Delta \Tc^{\mu\nu}_{L,\df}(P_n;q_1) \, , \nn 
\end{align}
which is exactly Eq.~\eqref{eq:Main_eq} as stated in Section~\ref{sec:Tdef}. Therefore, the desired $\Tc^{\mu\nu}_{\df}$ amplitude can be constrained from a linear combination of the previously introduced quantities and computable functions.

%%%%%%%%%%%%%%%%%%%%%%%%%%%%%%%%%%%%%%%%%%%%%%%%%%%%%%%%%%%%%%%%%%%%%%%%
%	Section :: Derivation
%%%%%%%%%%%%%%%%%%%%%%%%%%%%%%%%%%%%%%%%%%%%%%%%%%%%%%%%%%%%%%%%%%%%%%%%
\section{Derivation of finite-volume correction}
\label{sec:FVCorr}

In this section, we outline the derivation of the finite-volume correction for the long-range matrix element $\Tc_L$. The primary object considered is the three-point correlation function~\footnote{We follow the convention of Ref.~\cite{Briceno:2019opb} that operators are associated with a factor of $i$.}
\begin{align}
	\label{eq:Cggpipi}
	\Cc_L^{\mu\nu}(P;q_1)
	\equiv
	-\int_L \diff^4 x \int_L \diff^4 y \, e^{-i q_1 \cdot x} \, e^{i P \cdot y} \, \bra{\Omega}   \mathrm{T} \{\mathcal{O}(y) \, \Jc^{\mu} (x) \, \Jc^{\nu} (0) \} \ket{\Omega} \, ,
\end{align}
where we work in Minkowski spacetime, and $\mathcal{O}$ is an interpolating operator with quantum numbers of the two-pion system. The spatial integrations are over the finite-cubic-volume, while the temporal component is over an infinite range. To derive the finite-volume correction, we first identify the spectral decomposition of Eq.~\eqref{eq:Cggpipi} in terms of the matrix element desired, $\Tc_L^{\mu\nu}$. Then, we analyze the same correlation function using some generic relativistic effective field theory to generate an all-orders representation. By using the procedure introduced in Ref.~\cite{Kim:2005gf}, we then isolate all finite-volume contributions which have power-law dependence on $L$, which as we show, arise from on-shell two-particle intermediate states. Identifying these contributions allows us to write the desired matrix element in terms of its infinite-volume counterpart and a correction that can be written in terms of known amplitudes and geometric functions.

As we show in Section~\ref{sec:CLmunu}, the correlation function $\Cc_L^{\mu\nu}$ has pole singularities when the final two-particle states have energies that coincide with those allowed in a finite volume. These are well-known to be described by the two-particle quantization condition~\cite{Luscher:1986pf,Rummukainen:1995vs,Kim:2005gf,Fu:2011xz,He:2005ey,Briceno:2012yi,Hansen:2012tf,Briceno:2014oea}
\begin{align}
	\label{eq:QC}
	\det\left[F^{-1}(P,L) + \Mc(P^2)\right]_{P = P_n} = 0 \, ,
\end{align}
where $F$ is a well-known geometric function, defined in Appendix~\ref{app:FGL}, $\Mc$ is the infinite-volume two-particle scattering amplitude shown in Fig.~\ref{fig:iM_iH_iW}, and the determinant is over the space of angular momenta. The residue of $\Cc_L^{\mu\nu}$ at these poles can be related to the finite-volume matrix element $\Tc_L^{\mu\nu}$ through a multiplicative factor controlled by the two-particle correlation function. Therefore, to isolate the matrix element, we first summarize the analysis as outlined in Ref.~\cite{Briceno:2014uqa} to isolate finite-volume corrections of the two-point correlation function, and identify the scaling behavior near the energies of the finite-volume spectrum.

%%%%%%%%%%%%%%%%%%%%%%%%%%%%%%%%%%%%%%%%%%%%%%%%%%%%%%%%%%%%%%%%%%%%%%%%
%	Section :: Two-point
%%%%%%%%%%%%%%%%%%%%%%%%%%%%%%%%%%%%%%%%%%%%%%%%%%%%%%%%%%%%%%%%%%%%%%%%
\subsection{Two-point correlation function and residues}
\label{sec:CL}

%%%%%%%%%%%%%%%%%%
%	figure
%%%%%%%%%%%%%%%%%%
\begin{figure}[t]
	\begin{center}
	\includegraphics[width=1\textwidth]{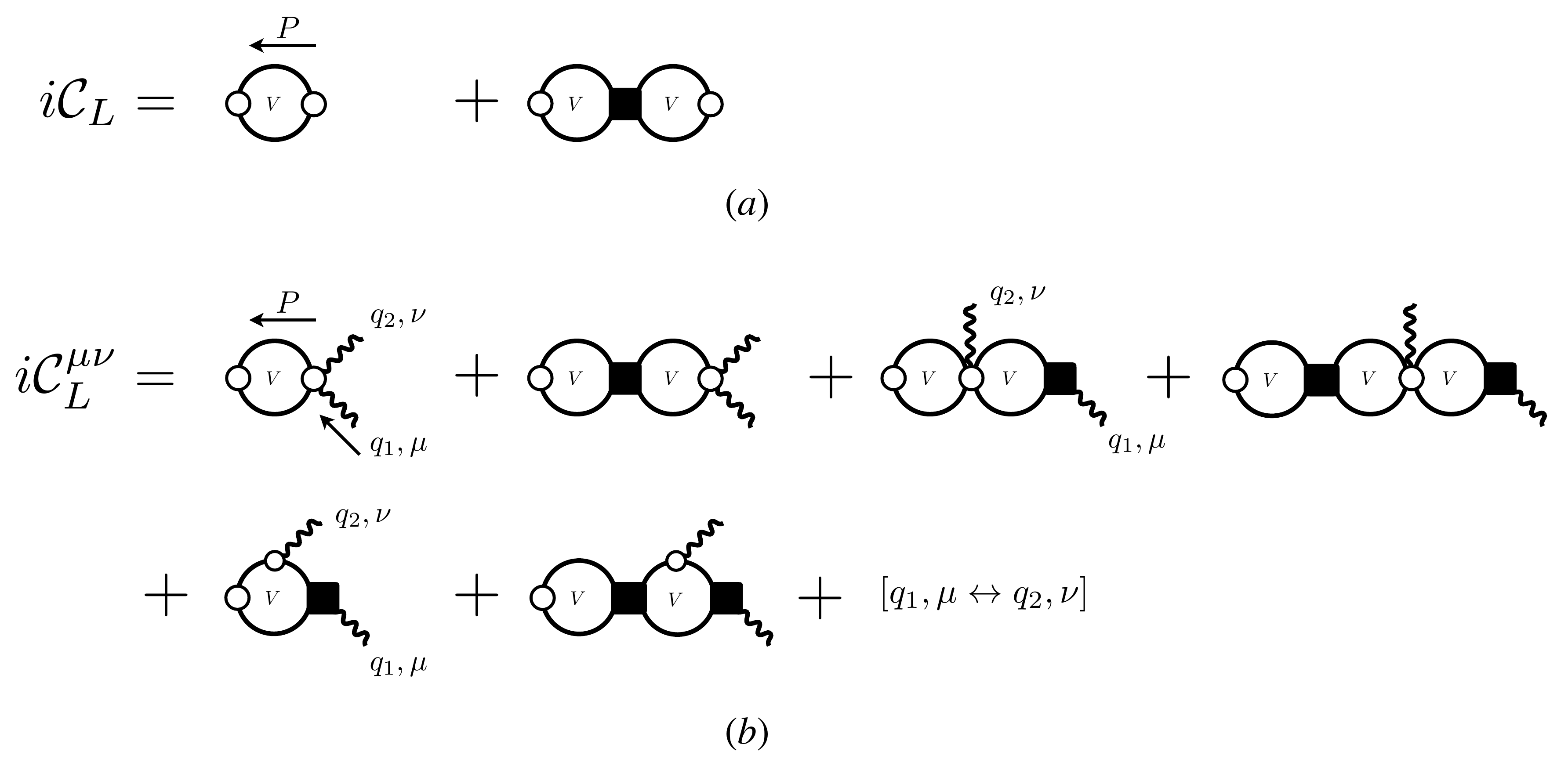}
	\caption{
	(a) Shown is the all-orders expression for the finite-volume two-point correlation function defined in Eq.~\eqref{eq:Cpipi}.
	(b) Explicitly shown are the \emph{direct} diagrams contributing to the three-point correlation function coupling $\pi\pi$ (solid lines) to two electromagnetic currents, cf. Eq.~\eqref{eq:Cggpipi}. The crossed-channel contributions are left implicit by the ``$\left[q_1,\mu\leftrightarrow q_2,\nu\right] $" symbol. All objects appearing were previously shown in Figs.~\ref{fig:iT} and \ref{fig:iM_iH_iW}, except for the black squares which are finite-volume analogues of their corresponding infinite volume amplitude, and the open circle coupling the vacuum to the intermediate two-particle states. The former is  defined in Fig.~\ref{fig:iML_iHL}, while the latter is understood as the undressed overlap between the two-particle operator and the finite-volume two-particle state.
	}
	\label{fig:iCLmunu}
	\end{center}
\end{figure}
%%%%%%%%%%%%%%%%%%
%	figure
%%%%%%%%%%%%%%%%%%

In order to analyze the three-point function Eq.~\eqref{eq:Cggpipi}, we first review the spectral content and finite-volume corrections for the two-point function of the two-particle interpolation operator $\mathcal{O}$. We follow the procedure outlined in Ref.~\cite{Kim:2005gf,Briceno:2014oea} and point the reader there for details. We evaluate the finite-volume Minkowski-signature two-point correlation function $i\Cc_L$ defined as
\begin{align}
	\label{eq:Cpipi}
	\Cc_L(P) \equiv  i\int_L \diff^4 x \, e^{i P \cdot x - \epsilon |x^0|}\, \bra{\Omega} \mathrm{T} \{\mathcal{O}(x) \, \mathcal{O}^\dagger(0) \} \ket{\Omega}  \, ,
\end{align}
where it is assumed $\epsilon \to 0^{+}$ after integration. We construct the spectral decomposition by inserting a complete set of finite-volume states $\ket{P_n,L}$ between the interpolating operators for both time orderings. Identifying $\mathcal{O}$ as a Heisenberg operator, we evaluate the spatial and temporal integrals to obtain
\begin{align}
	\label{eq:Cspecpipi_comp}
	\Cc_L(P) = -L^3\sum_n \frac{Z_n Z^*_n}{E-E_n}
	+ L^3\sum_n \frac{Z_n^* Z_n}{E+E_n},
\end{align}
where $Z_n=\braket{\Omega| \mathcal{O}(0)|P_n,L}$ is the overlap factor for the operator $\mathcal{O}$ on the state $n$. For notational convenience, we suppress the dependence of $Z_n$ on $L$ and $\mathbf{P}$. Ultimately, we are interested in the correlation functions near a given finite-volume state $n$ with a given energy $E_n$. As can be seen from Eq.~\eqref{eq:Cspecpipi_comp}, in the vicinity of the given eigenstate $\ket{P_n,L}$ for some fixed $L$ and $\mathbf{P}$, the correlation function behaves like
\begin{align}
	\label{eq:Cspecpipi}
	\Cc_L(P) \sim -L^3 \frac{Z_n Z^*_n}{E-E_n} \, ,
\end{align}
as $E \sim E_n$. Thus, the correlation diverges as a pole singularity in the energy near the $n$th eigenstate.

\begin{figure}[t]
\begin{center}
\includegraphics[width=1\textwidth]{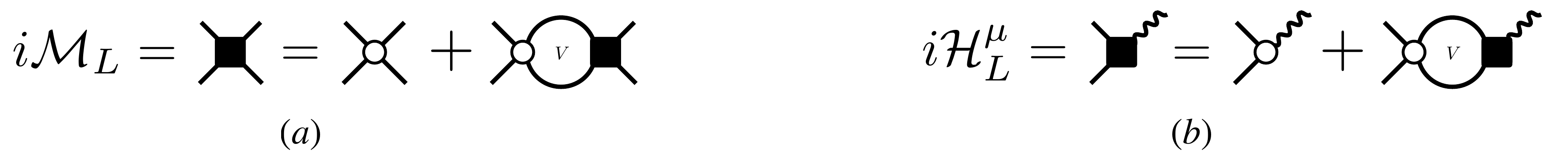}
\caption{ Shown is the diagrammatic definition of the finite-volume analogues of the (a) $\pi\pi\to\pi\pi$, (b) $\gamma^\star\to\pi\pi$ amplitudes. The ``$V$" symbol emphasizes that the loops are being evaluated in a finite volume. The open circles are the same kernels appearing in the definition of the infinite-volume amplitudes, shown in Fig.~\ref{fig:iM_iH_iW}.}
\label{fig:iML_iHL}
\end{center}
\end{figure}

We now express the same correlation function to all orders in our generic relativistic quantum field theory. Summing to all-orders, we find that $i\Cc_L$ can be written diagrammatically as shown in Fig.~\ref{fig:iCLmunu}(a), where the vacuum-to-two-particle kernels, which we denote as $i\sigma$, are represented by white open circles. The black square is a finite-volume analogue of the off-shell $2\to 2$  amplitude $\Mc_L$, defined diagrammatically in Fig.~\ref{fig:iML_iHL}(a). Note that the loops shown in these figures are not integrations over the temporal components of momenta, but finite-volume sums over the quantized spatial components of momenta, which is indicated by the ``$V$'' in the figures.

%%%%%%%%%%%%%%%%%%
%	figure
%%%%%%%%%%%%%%%%%%
%%%%%%%%%%%%%%%%%%
%	figure
%%%%%%%%%%%%%%%%%%

\begin{figure}[t]
	\begin{center}
	\includegraphics[width=.95\textwidth]{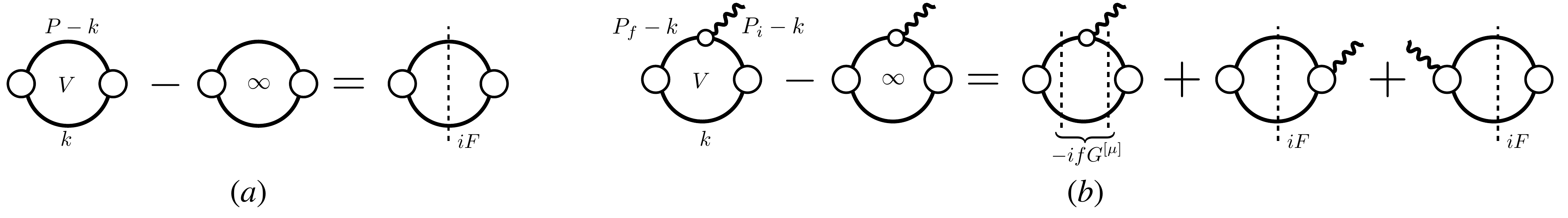}
	\caption{(a) Shown is a diagrammatic representation of Eq.~\eqref{eq:on_shell_loop}, which relates the finite-volume loops, infinite-volume loops, and the finite-volume $F$ function defined in Eq~\eqref{eq:FV_F}. (b) Shows the diagrammatic definition of the finite-volume $G^{[\mu]}$ function as given in Eq.~\eqref{eq:Gmu_tot} and subsequently in Eq.~\eqref{eq:FV_G}.
	}
	\label{fig:Loop}
	\end{center}
\end{figure}

To extract the finite-volume correction to the infinite-volume correlation function $i\Cc_{\infty}$, we note that power-law finite-volume effects are understood to arise from on-shell intermediate states. Therefore, we do not need to make assumptions about the explicit forms of the short-distance kernels or propagators. Instead, all that is needed is to identify that in the vicinity of a given particle going on-shell, the propagator behaves as $i\Delta \sim iD$ as $k^2 \sim m^2$, where $iD$ is defined in Eq.~\eqref{eq:D_bareprop}, and we can project the kernels to their on-shell counterpart by expanding about this point. Using this fact, we systematically replace each finite-volume loop with an infinite-volume loop plus the difference between the two. The difference between the finite- and infinite-volume loops, shown in Fig.~\ref{fig:Loop}(a), can be shown to satisfy the relation~\cite{Kim:2005gf,Briceno:2014oea}
\begin{equation}
	\label{eq:on_shell_loop}
	\xi \left[\frac{1}{L^3}\sum_{\textbf{k}} - \int \!\frac{\diff^3 \mathbf{k}}{(2\pi)^3}\right]  \int\frac{\diff k^0}{2\pi} \, i\sigma(P,k) \, i \Delta(k)\, i \Delta(P-k) \, i\sigma^\dag(P,k)= i\sigma_{\rm on}(P) \, \cdot iF(P,L) \, \cdot i\sigma^\dag_{\rm on}(P) \, ,
\end{equation}
where $i\sigma^\dag$ is the overlap between the creation operator and the two-particle state before dressing to all orders, $i\Delta$ denotes the fully dressed propagator for an individual particle, $\xi$ is a symmetry factor that is equal to $1/2$ if the particles are identical or $1$ otherwise, and $F$ is the known finite-volume function defined in~\cref{app:FGL}. The subscript ``$\rm on$" emphasizes that the functions have been placed on-shell and partial-wave projected. The $\cdot$ represents matrix multiplication in angular momentum space. Equation~\eqref{eq:on_shell_loop} holds up to $\mathcal{O}(e^{-m L})$ corrections, which we ignore throughout this work as we assume $mL \gg 1$.

Following this procedure to all orders, for identical two-pion initial and final states, one finds that the two-point correlation function can be decomposed as~\cite{Kim:2005gf}~\footnote{Note, the factor of $i$ appearing in the left-hand side of Eq.~\eqref{eq:CcL} is consistent with Eq.~\eqref{eq:Cpipi}. It is easy to convince oneself that the diagrammatic representations can be identified with $i\Cc_L$ if one has a factor of $i$ for each vertex shown in the diagrams, including vertices associated with interpolating operators and current insertions.}
\begin{align}
	\label{eq:CcL}
	i\Cc_L(P) = i\Cc_\infty (P) + i\Ac(P) \cdot i\Fc_{L}(P,L) \cdot i\Ac^\dag(P) \, ,
\end{align}
where $\Ac$ is the fully-dressed vacuum to the $\pi\pi$ amplitude that encodes the rescattering physics applied to $i\sigma_{\mathrm{on}}$, and $\Fc_{L}$ is defined as
\begin{align}
	\label{eq:f2l}
	\Fc_{L}(P,L)\equiv  \frac{1}{F^{-1}(P,L)+\Mc(P^2)} \, .
\end{align}
The function $\Fc_L$ is the result of summing the geometric series that arises when we use the loop difference identity Eq.~\eqref{eq:on_shell_loop} to all orders, as discussed in Refs.~\cite{Kim:2005gf,Briceno:2014oea}.

Near the vicinity of the $n$th eigenstate with energy $E_n$, the correlation function diverges as $E \sim E_n$. We identify that these poles must occur when $\det\left[\Fc_L^{-1}(P_n,L)\right] = 0$ since the remaining functions are infinite-volume quantities, which gives the L\"uscher quantization condition as stated in Eq.~\eqref{eq:QC}. For $E \sim E_n$, we can rewrite $\Fc_{L}$ in terms of its residue near the $n$th state
\begin{align}
	\label{eq:F2L_pole}
	\Fc_{L}(P,L) \sim \frac{\widetilde{\Rc}_n}{2E_n(E-E_n)} \, ,
\end{align}
where the residue $\widetilde{\Rc}_n$ is the generalized Lellouch-L\"uscher factor as presented in Ref.~\cite{Briceno:2014uqa}.~\footnote{Here we choose a more convenient normalization for the residue as compared to Ref.~\cite{Briceno:2014uqa}, which was denoted by $\Rc$. The connection between normalizations is $\widetilde{\Rc}_n = 2E_n \, \Rc_n$.} Note that the residue implicitly depends on $L$ and $\mathbf{P}$. For practical applications, it is convenient to write the residue as an eigendecomposition of the matrix $F^{-1}+\Mc$~\cite{Briceno:2021xlc},
\begin{align}
	\label{eq:Rc}
	\widetilde{\Rc}_n=\left(-\frac{2 E^\star_n}{{\mu^{\star}_0}'}\right) \, 
	\Mc^{-1}\, \mathbf{w}_0 \otimes
	\mathbf{w}^{\top}_0 \,
	\Mc^{-1} \, ,
\end{align}
where $E^\star_n = \sqrt{E_n^2 - \mathbf{P}^2}$ is the CM energy of the $n$th state, $\mu^{\star }_0$ is the vanishing eigenvalue of $F^{-1}+\Mc$ at this finite-volume energy, $\mathbf{w}_0$ is the corresponding eigenvector, and 
\begin{align}
	{\mu^{\star}_0}'\equiv \frac{\diff \mu^{\star }_0}{\diff E^{\star }}\bigg|_{E^{\star} = E^{\star }_n} \, .
\end{align}
Note that $\Mc^{-1}$ in Eq.~\eqref{eq:Rc} is evaluated at $P_n$.

%%%%%%%%%%%%%%%%%%
%	figure
%%%%%%%%%%%%%%%%%%
%%%%%%%%%%%%%%%%%%
%	figure
%%%%%%%%%%%%%%%%%%

We simplify the expression further by introducing a compact notation for the finite-volume residue
\begin{align}
	\label{eq:Rvec}
	\rrvec\equiv \sqrt{-\frac{2 E^\star_n}{{\mu^{\star}_0}'}} \,
	\mathbf{w}^{\top}_0 \, 
	\Mc^{-1} \, ,
\end{align}
where the arrow is to remind the reader that this is a vector in the space of the degrees of freedom of the scattering systems, which in this case we assume to only be angular momentum. With this definition, Eq.~\eqref{eq:Rc} can be rewritten as
\begin{align}
	\label{eq:Rc_vf}
	\widetilde{\Rc}_n=\lrvec \, \otimes
	\rrvec \, .
\end{align}
This factorized residue allows us to write the correlation function near the pole as
\begin{align}
	i\Cc_L 
	&\sim i\Ac(P_n) \cdot \frac{i\bigg(\lrvec \, \otimes
	\rrvec\bigg)}{2E_n(E-E_n)}
	\cdot i\Ac^\dag(P_n)
	 =  -L^3 \frac{iZ_n Z^*_n}{E-E_n} \, ,
\end{align}
where in the last equality we have used the spectral decomposition from~\cref{eq:Cspecpipi}. Comparing the residue of the spectral decomposition to that derived from the all-orders representation allows us to relate the overlap factors $Z_n$ involving the creation operations to the infinite-volume vacuum-to-two-pion amplitude $\Ac$,~\footnote{Note that, although in general unphysical, we can write an on-shell representation for $\Ac$, similar to $\Hc^\mu$ in Eq.~\eqref{eq:H_on}, in which one would arrive at the two-particle scattering amplitude $\Mc$ multiplying some smooth function. From this, we find that $\Mc$ cancels $\Mc^{-1}$ in the definition of $\lrvec$ given similarly to Eq.~\eqref{eq:Rvec}.}
\begin{align}
	\label{eq:Zn} 
	Z_n\equiv
	\Ac(P_n) \cdot\frac{\lrvec}{\sqrt{2 E_n L^3}} \, . 
\end{align}
As we show in the following section, this relation allows us to isolate the matrix element $\Tc_L$ from the three-point function.

%%%%%%%%%%%%%%%%%%%%%%%%%%%%%%%%%%%%%%%%%%%%%%%%%%%%%%%%%%%%%%%%%%%%%%%%
%	Section :: Three-point
%%%%%%%%%%%%%%%%%%%%%%%%%%%%%%%%%%%%%%%%%%%%%%%%%%%%%%%%%%%%%%%%%%%%%%%%
\subsection{Three-point correlation function}
\label{sec:CLmunu}

We proceed to perform a similar investigation on the three-point correlation function of interest, defined in~\cref{eq:Cggpipi}. Once again we start by briefly sketching its spectral decomposition, obtained by inserting complete sets of states $\ket{P_n,L}$ between the interpolating operators. Since we have three operators inside the time-ordering operator, the spectral decomposition of the three-point correlation function is slightly more complicated. For simplicity, we focus our attention on those terms that lead to the desired simple-pole contribution in the vicinity of $E\sim E_n$. These come from those associated with $y^0>x^0>0$ and $y^0>0>x^0$. To simplify the algebra, let us define $z^0 =\max(x^0, 0)$. Then, it is easy to see that the pole contribution we are after can be written as
\begin{align}
	\Cc_L^{\mu\nu}(P;q_1) &\equiv  -\int_L\!\diff^4 x \int_L\!\diff^4 y \, e^{-i q_1 \cdot x-\epsilon |x^0|}\, e^{i P \cdot y-\epsilon |y^0|} \, \bra{\Omega}  \mathrm{T} \{
	\mathcal{O}(y)\Jc^{\mu} (x) \Jc^{\nu} (0) \} \ket{\Omega}  \, ,
	\, \nn \\[5pt]
	& \sim -L^3 \frac{Z_n}{E-E_n} \, i \int_L \diff^4 x \, e^{-i q_1 \cdot x-\epsilon |x^0|}\,e^{i (E-E_n) \cdot z^0 } \, \bra{P_n,L} \mathrm{T} \{\Jc^{\mu} (x) \Jc^{\nu} (0) \} \ket{\Omega} \, .
\end{align}
In the second line, we consider $E \sim E_n$ and ignore time-orderings that do not give the pole structure we are after. One sees that in the vicinity of the $n$th pole, the $z^0$-dependent phase vanishes, and one arrives at the simplified function
\begin{align}
	\label{eq:Cspecggpipi}
	\Cc_L^{\mu\nu}(P;q_1)
	&\sim
	-L^3 \frac{Z_n}{E-E_n}  \, i \int_L \diff^4 x \, e^{-i q_1 \cdot x-\epsilon |x^0|} \, \bra{P_n,L} \mathrm{T} \{\Jc^{\mu} (x) \Jc^{\nu} (0) \} \ket{\Omega} \, ,
	\nn\\[5pt]
	& \equiv -L^3
	\frac{Z_n}{E-E_n} \, 
	\Tc_L^{\mu\nu}(P_n;q_1)
	, 
\end{align}
where $\Tc_L^{\mu\nu}$ is the desired quantity introduced in Eq.~\eqref{eq:TLrel}. 

%%%%%%%%%%%%%%%%%%
%	figure
%%%%%%%%%%%%%%%%%%
\begin{figure}[t]
\begin{center}
\includegraphics[width=1\textwidth]{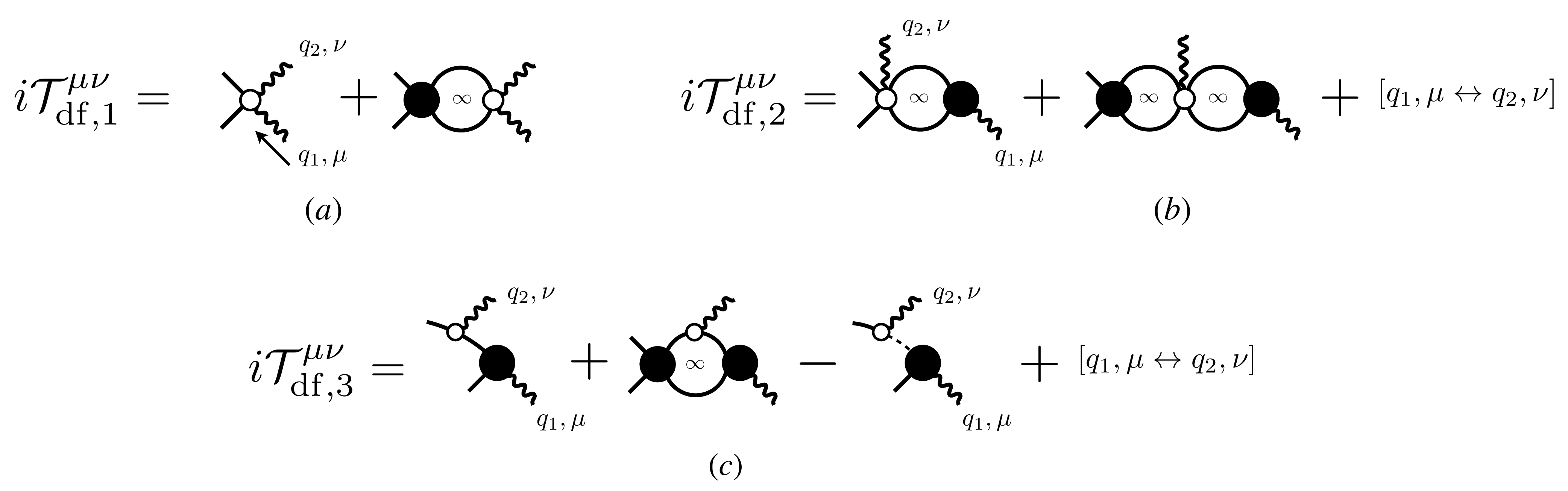}
\caption{Shown are the three contributions to the all-orders definition of $\Tc^{\mu\nu}_{\df}$. Each term is distinguished by the types of kernels coupling the currents to the particles.
}
\label{fig:iTdf}
\end{center}
\end{figure}
%%%%%%%%%%%%%%%%%%
%	figure
%%%%%%%%%%%%%%%%%%

We can now proceed to study the all-orders representation of this correlation function, which is diagrammatically depicted in Fig.~\ref{fig:iCLmunu}(b). It is convenient to break up the correlation function into sectors of topologically similar diagrams. We write $i\Cc_L^{\mu\nu}$ as the sum of three contributions,
\begin{align}
	\label{eq:Cmunu_break}
	i\Cc_L^{\mu\nu}(P;q_1) = i\Cc_{L,1}^{\mu\nu}(P;q_1) + i\Cc_{L,2}^{\mu\nu}(P;q_1) + i\Cc_{L,3}^{\mu\nu}(P;q_1) \, .
\end{align}
The first contribution, $i\Cc_{L,1}^{\mu\nu}$, is given by the first two terms of Fig.~\ref{fig:iCLmunu}(b) which involves currents coupling to two-particle states via a short-distance kernel. Next, $i\Cc_{L,2}^{\mu\nu}$ comes from the sum of the third and fourth terms of Fig.~\ref{fig:iCLmunu}(b) which contains a short-distance $\pi\pi \gamma^{\star} \to \pi\pi$ kernel and the finite-volume analogue of the $\gamma^{\star} \to \pi\pi$ amplitude which is diagrammatically defined in Fig.~\ref{fig:iML_iHL}(b). The last contribution, $i\Cc_{L,3}^{\mu\nu}$, constitutes the final two terms of Fig.~\ref{fig:iCLmunu}(b) which involves the $\pi \gamma^{\star} \to \pi$ kernel in the loops. Note that Eq.~\eqref{eq:Cmunu_break} also includes contributions from the $[q_1,\mu \leftrightarrow q_2,\nu]$ crossed channel diagrams in the $\Cc_{L,2}^{\mu\nu}$ and $\Cc_{L,3}^{\mu\nu}$ terms.~\footnote{In principle there can be additional terms that represent pure short-distance objects, such as two currents annihilating the vacuum. We ignore such terms as they do not contribute to the leading finite-volume behavior of the correlation function.}

We start by focusing our attention on $\Cc_{L,1}^{\mu\nu}$, the first two diagrams  in~\cref{fig:iCLmunu}(b) where the short-distance kernel involved currents coupling to two-pions. The classes of finite-volume diagrams are the same appearing in the two-point function considered in the previous section. Using the identity given in Fig.~\ref{fig:Loop}, one sees that all power-law finite-volume corrections are encoded in the previously introduced $F$ function. Therefore, we can immediately write the contribution from the first two diagrams as
\begin{align}
	\label{eq:Cmunu1}
	i\Cc_{L,1}^{\mu\nu}(P;q_1) = i\Cc_{\infty,1}^{\mu\nu}(P;q_1)+ i\Ac(P) \cdot i\Fc_{L}(P,L) \cdot i\Tc^{\mu\nu}_{\df, 1}(P;q_1) \, ,
\end{align}
where $\Tc_{\df, 1}^{\mu\nu}$ is a contribution to the divergence-free infinite-volume \ggpipi amplitude as defined in Section~\ref{sec:Tdef}. This object comes from summing a subset of the infinite-volume diagrams contributing to $\Tc_{\df}^{\mu\nu}$, which are shown explicitly in Fig.~\ref{fig:iTdf}(a).

\begin{figure}[t]
	\begin{center}
	\includegraphics[width=1\textwidth]{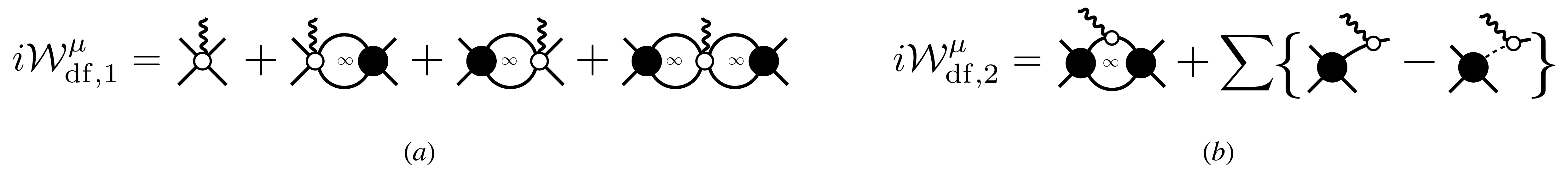}
	\caption{Shown are the two contributions to $\Wc^{\mu}_{\df}$ arising from (a) the two-body coupling to the current and (b) the one-body coupling to the current.
	}
	\label{fig:iWdf}
	\end{center}
\end{figure}

Next, we look at the diagrams which include current insertions separated by two-particle loops as defined in $i\Cc_{L,2}^{\mu\nu}$. When summed to all orders, these diagrams produce two different kinds of finite-volume corrections. The first one is proportional to another contribution to $\Tc_{\df}^{\mu\nu}$, which we label $\Tc_{\df, 2}^{\mu\nu}$ and is shown diagrammatically in Fig.~\ref{fig:iTdf}(b). The second class of finite-volume corrections is proportional to $\Wc_{\df,1}^{\mu}$ and $\Hc^{\nu}$ amplitudes. The $\gamma^{\star} \to \pi\pi$ transition amplitude, $\Hc^{\nu}$, was previously defined and shown diagrammatically in Fig.~\ref{fig:iM_iH_iW}(b). The $\Wc^\mu_{\df,1}$ function gives one contribution to the $\Wc^\mu_{\df}$ amplitude introduced in Section~\ref{sec:long_range_amplitude}. In particular, as shown Fig.~\ref{fig:iWdf}(a), this includes all diagrams that contain a $\pi\pi \gamma^{\star} \to \pi\pi$ short-distance kernel. Having introduced these objects, the second class of terms contributing to $i\Cc_L^{\mu\nu}$ can be written as,
\begin{align}
	\label{eq:Cmunu2}
	i\Cc_{L,2}^{\mu\nu}(P;q_1) = i\Cc_{\infty,2}^{\mu\nu}(P;q_1) & + i\Ac(P) \cdot i\Fc_{L}(P,L) \cdot i\Tc^{\mu\nu}_{\df, 2}(P;q_1) \, \nn \\[5pt]
	& + i\Ac(P) \cdot i\Fc_{L}(P,L) \cdot i \Wc^\nu_{\df,1}(P;q_1) \cdot i\Fc_{L}(q_1,L) \cdot i\Hc^{\mu}(q_1)  + \left[q_1,\mu\leftrightarrow q_2,\nu\right] \,,
\end{align}
where $\left[q_1,\mu\leftrightarrow q_2,\nu\right]$ affects only the third term on the right-hand-side.

%%%%%%%%%%%%%%%%%%
%	figure
%%%%%%%%%%%%%%%%%%
%%%%%%%%%%%%%%%%%%
%	figure
%%%%%%%%%%%%%%%%%%

Finally, the last set of diagrams contributing to $\Cc_L^{\mu\nu}$ involves a new class of finite-volume functions, produced by the current coupling to one-particle states that can go on-shell. These are the finite-volume analogue of the triangle functions~\cite{Briceno:2015tza, Baroni:2018iau}. The difference between the finite- and infinite-volume triangle diagrams is shown in Fig.~\ref{fig:Loop}(b). Following a similar analysis, as was done in Eq.~\eqref{eq:on_shell_loop} to the simple s-channel loop, one can show that the difference is given by three terms. Two of these terms depend solely on the $F$ function, while the third new term depends on the finite-volume triangle function. Since here we consider the insertion of a conserved vector current, this geometric function has an explicit dependence on the Lorentz index of the current. In Ref.~\cite{Baroni:2018iau}, this function was shown to be written in terms of two irreducible geometric functions. Labeling this geometric function as $G^{[\nu]}$, we can write it as
\begin{align}
	\label{eq:Gmu_tot}
	G^{[\nu]}(P_f,P_i,L) \equiv (P_f+P_i)^\nu \, G(P_f,P_i,L) -2 G^{\nu} (P_f,P_i,L),
\end{align}
where $P_i$ and $P_f$ denote generic incoming and outgoing momenta flowing through the function. The $G$ and $G^{\nu}$ functions are closely related, with $G^{\nu}$ depending explicitly on a $\nu$th-component of the internal particle four-momentum $k^{\nu}$ that does not couple directly to the external current. In~\cref{app:FGL}, we give explicit expressions for these functions, cf. Eq.~\eqref{eq:FV_G}.

The resulting contribution to the correlation function, which we labeled $\Cc_{L,3}^{\mu\nu}$, depend on the remaining pieces of the $\Tc_{\df}^{\mu\nu}$ and $\Wc^\nu_{\df}$ amplitudes, which we denote as $\Tc_{\df,3}^{\mu\nu}$ and $\Wc^\nu_{\df,2}$, respectively. These include the triangle singularities and are depicted in Fig.~\ref{fig:iTdf}(c) and Fig.~\ref{fig:iWdf}(b), respectively. The remaining finite-volume correction to $\Cc_{L,3}^{\mu\nu}$ depends on $G^{[\nu]}$, $\Hc^\mu$, and $\Mc$. Altogether, $\Cc_{L,3}^{\mu\nu}$ can be written as 
\begin{align}
	\label{eq:Cmunu3}
	i\Cc_{L,3}^{\mu\nu}(P;q_1) = i\Cc_{\infty,3}^{\mu\nu}(P;q_1)
	& +i\Ac(P) \cdot i\Fc_{L}(P,L) \cdot i\Tc^{\mu\nu}_{\df, 3}(P;q_1) \nn \\[5pt]
	& + 
	i\Ac(P)
	\cdot i\Fc_{L}(P,L) \cdot i \Wc^\nu_{\df,2}(P;q_1) \cdot 
	i\Fc_{L}(q_1,L) \cdot
	i\Hc^{\mu}(q_1) 
	\nn\\[5pt]
	& 
	-i\Ac(P)
	\cdot i\Fc_{L}(P,L) \cdot
	i\Mc(P^2) \cdot 
	if G^{[\nu]}(P,q_1,L)
	 \cdot 
	[ 1+i\Mc(q_1^2) \cdot i\Fc_{L}(q_1,L)  ]
	\cdot i\Hc^{\mu}(q_1) 
	\, \nn \\[5pt]
	&  + \left[q_1,\mu\leftrightarrow q_2,\nu\right],
\end{align}
where $f = f(Q_2^2)$ is the on-shell form factor of pion and $\left[q_1,\mu\leftrightarrow q_2,\nu\right]$ affect only the third and fourth terms on the right-hand-side.

Combining Eqs.~\eqref{eq:Cmunu1}, \eqref{eq:Cmunu2}, and \eqref{eq:Cmunu3} into Eq.~\eqref{eq:Cmunu_break}, we arrive at an analytic expression to the finite-volume correction to $\Cc_L^{\mu\nu}$, 
\begin{align}
    \label{eq:corr_delta_CL}
	\Delta i\Cc_L^{\mu\nu}(P;q_1)
	& \equiv i\Cc_L^{\mu\nu}(P;q_1) - i\Cc_\infty^{\mu\nu}(P;q_1)\, \nn \\[5pt]
	& = i\Ac(P) \cdot i\Fc_{L}(P,L) \cdot \bigg( \, i\Tc^{\mu\nu}_{\df}(P;q_1)- i\Mc(P) \cdot 
	if G^{[\nu]}(P,q_1,L)\cdot i\Hc^{\mu}(q_1) \, \bigg)
	\nn\\[5pt]
	& \qquad +
	i\Ac(P)
	\cdot i\Fc_{L}(P,L) \cdot
	\bigg( 
	i \Wc^\nu_{L,\df}(P;q_1) \cdot 
	i\Fc_{L}(q_1,L) \cdot i\Hc^{\mu}(q_1) \bigg) +\left[q_1,\mu\leftrightarrow q_2,\nu\right] \, ,
\end{align}
where we followed Eq. (97) of Ref.~\cite{Briceno:2015tza} and introduced
\begin{align}
	i \Wc^\nu_{L,\df}(P;q_1) \equiv \bigg( 
	i \Wc^\nu_{\df}(P;q_1)  
	- i\Mc(P) \cdot 
	i f G^{[\nu]}(P,q_1,L) \cdot i\Mc(q_1^2)
	\bigg) \, .
\end{align}
Note that $\left[q_1,\mu\leftrightarrow q_2,\nu\right]$ in Eq.~\eqref{eq:corr_delta_CL} applies to all terms except the one which contains $\Tc_{\df}$. Using Eq.~\eqref{eq:F2L_pole} for the behavior of $\Fc_{L}$ near the L\"uscher poles, Eq.~\eqref{eq:Rc_vf} for the expression for the residue, Eq.~\eqref{eq:Zn} for $Z_n$, and the spectral representation for $\Cc_{L}^{\mu\nu}$ given in Eq.~\eqref{eq:Cspecggpipi}, we find that the scaling behavior at the $n$th eigenstate gives the matrix element
\begin{align}
	\Tc_L(P_n;q_1)
	\equiv
	\frac{\rrvec}{\sqrt{2 E_n L^3}}\cdot
	\bigg(
	\Tc^{\mu\nu}_{\df}(P_n;q_1)
	-\Delta \Tc^{\mu\nu}_{L,\df}(P_n;q_1)
	\bigg) \, ,
\end{align}
where we define the correction $\Delta \Tc^{\mu\nu}_{L,\df}$ as
\begin{align}
	\label{eq:deltaTLdf}
	\Delta \Tc^{\mu\nu}_{L,\df}(P;q_1)
	=
	\bigg(\Wc^\nu_{L,\df}(P;q_1) \cdot 
	\Fc_{L}(q_1,L) -
	\Mc(s) \cdot 
	f(q^2_2)\, G^{[\nu]}(P,q_1,L)
	\bigg)
	\cdot \Hc^{\mu}(q_1)
	+\left[q_1,\mu\leftrightarrow q_2,\nu\right].
\end{align}
This is the necessary quantity to correct the finite-volume effects of the linear combination of correlators, as defined in Eq.~\eqref{eq:Main_eq}. In particular, what is needed is the product $\rrvec\cdot \Delta \Tc^{\mu\nu}_{L,\df}(P_n;q_1) \, / \, \sqrt{2 E_n L^3}$, which is the final and main result of this section. 
 
As a first theoretical test of our result, we provide a simple consistency check of the formalism in Appendix~\ref{app:deep_bs} when the two-pion system forms a deep bound state and no intermediate states can go on shell. There, we show that Eq.~\eqref{eq:deltaTLdf} coincides with the infinite-volume matrix elements up to the necessary relativistic normalization for a single-particle state.

%%%%%%%%%%%%%%%%%%%%%%%%%%%%%%%%%%%%%%%%%%%%%%%%%%%%%%%%%%%%%%%%%%%%%%%%
%	Section :: Conclusions
%%%%%%%%%%%%%%%%%%%%%%%%%%%%%%%%%%%%%%%%%%%%%%%%%%%%%%%%%%%%%%%%%%%%%%%%
\section{Conclusions}
\label{sec:conclusions}
In this work, we have derived a framework, defined by our main result in Eq.~\eqref{eq:Main_eq}, which allows us to access the \ggpipi amplitude from quantities that can be constrained directly via lattice QCD. The formalism outlined here relates three-point correlation functions of currents displaced in time with the desired long-range amplitude. This approach is model-independent and follows similar steps to previous formalism derived for a simpler class of amplitudes of the form $1+\Jc \to 2\to 1 + \Jc$~\cite{Briceno:2019opb}.

The desired amplitude can be decomposed in terms of different on-shell quantities describing possible physical subprocesses, namely $\pi \pi \to \pi\pi$, $\pi \gamma^\star \to \pi$, and $\gamma^\star \to\pi\pi$. The analytic structure of the amplitudes for the subprocesses is well-understood ~\cite{Briceno:2020vgp,Sherman:2022tco}, and it is now well-known how these may be constrained via lattice QCD.

Equation~\eqref{eq:Main_eq} removes all power-law finite-volume artifacts associated with the intermediate and final $\pi\pi$ states. It is exact up to suppressed effects that scale as $\mathcal{O}(e^{-m_\pi L})$. This formalism provides a key step towards extracting more complicated phenomena from lattice QCD correlation functions. For example, extensions of this work will provide a framework for determining long-range nucleon-nucleon processes from lattice QCD, including neutrino-less double-beta transitions ~\cite{Cirigliano:2022oqy}.~\footnote{For ongoing formal developments in this direction using a non-relativistic effective field theory approach, we point the reader to Refs.~\cite{Davoudi:2020xdv,Davoudi:2020gxs,Davoudi:2021noh}.} Furthermore, given the tremendous progress in extending finite-volume formalisms for three-particle systems~\cite{Hansen:2015zga,Mai:2017bge,Briceno:2012rv,Hammer:2017kms} and the recent lattice calculation of such processes~\cite{Blanton:2019vdk,Hansen:2020otl,Mai:2021nul,Blanton:2021eyf}, it is not hard to imagine extending the kinematic region of applicability of this work to energies above three-particle thresholds where, for example, $\gamma^\star\to 3\pi+\gamma^\star\to 2\pi$ can lead to a new class of power-law effects.

%%%%%%%%%%%%%%%%%%%%%%%%%%%%%%%%%%%%
%	Acknowledgements
%%%%%%%%%%%%%%%%%%%%%%%%%%%%%%%%%%%%

 \section{Acknowledgments}
This work is supported in part by USDOE grant No. DE-AC05-06OR23177, 
under which Jefferson Science Associates, LLC, manages and operates Jefferson Lab.
RAB and AWJ also acknowledge support from the USDOE Early Career award, contract DE-SC0019229. JVG is supported by the Jefferson Lab LDRD project LD2117. AR acknowledges the financial support of the U.S. Department of Energy contract DE-SC0018416 at the College of William \& Mary.

%%%%%%%%%%%%%%%%%%%%%%%%%%%%%%%%%%%%%%%%%%%%%%%%%%%%%%%%%%%%%%%%%%%%%%%%
%	Appendices
%%%%%%%%%%%%%%%%%%%%%%%%%%%%%%%%%%%%%%%%%%%%%%%%%%%%%%%%%%%%%%%%%%%%%%%%
\appendix

%%%%%%%%%%%%%%%%%%%%%%%%%%%%%%%%%%%%
%	App :: Lorentz decomposition
%%%%%%%%%%%%%%%%%%%%%%%%%%%%%%%%%%%%
\section{$\Tc^{\mu\nu}$ Lorentz decomposition}
\label{app:decomp}
In this section, we discuss the Lorentz decomposition for the most immediately relevant case where the final $\pi\pi$ state has been projected onto the $J^{PC}=0^{++}$ channel. For this channel, only the $\ell =0$ component of $\Tc^{\mu\nu}$ and $\Bc^{\mu\nu}_{20}$ contribute. Because of the quantum numbers of the current, the only allowed long-range processes must involve intermediate $\pi\pi$ states that have the quantum number of the $\rho(770)$, $1^{--}$. This means that the only contributing pieces for $\Hc^{\mu}_{\ell m}$ would be $\Hc^{\mu}_{1m}$, where $m$ can run over $-1,0$, and $1$.

The $\gamma^\star\gamma^\star \to  (\pi\pi)_{0^{++}}$ amplitude can in general be written in terms of two $s$-dependent form factors. One can show this in at least two ways. Arguably the simplest is to first construct all possible kinematic structures that may couple a vector-vector state to scalar one. In total, there are five such tensors leading to five unknown form factors. By then imposing the Ward-Takahashi identities, this is reduced to two.

Alternatively, one can begin by identifying the general  $\gamma^\star\gamma^\star \to  \pi\pi $ amplitude and imposing the  Ward-Takahashi identities. Following this procedure, one finds that the amplitude can be written in terms of only five different $s$-, and $t$-dependent form factors~\cite{Colangelo:2014dfa,Colangelo:2015ama,Danilkin:2019opj}. One can then proceed to project the subsequent amplitude to the desired angular momentum.

Given that the later amplitudes are most commonly used in the literature, here we explain how these are projected to $0^{++}$ and are related to the desired $\Tc_\df$. We begin by rewriting the known Lorentz decomposition of the amplitude~\cite{Colangelo:2014dfa,Colangelo:2015ama,Danilkin:2019opj},
\begin{align}
	\label{eq:genTdecomp}
    \Tc^{\mu\nu}(P,\hat{\mathbf{p}}^{\star};q_1) = \sum_{j=1}^5 h_j(s,t) K_j^{\mu \nu},
\end{align}
where $s=(q_1+q_2)^2,\,t=(q_1-p_1)^2$ are the Mandelstam variables, $K_j^{\mu \nu}$ are known kinematic tensors, and $ h_j(s,t)$ are unknown form factors. The tensor structures can be conveniently chosen as~\cite{Danilkin:2019opj}
\begin{align}
\label{eq:Ktensors}
    K_1^{\mu \nu} & = q_1^\nu q_2^\mu - (q_1\cdot q_2)g^{\mu \nu} ,\\
    K_2^{\mu \nu} & = \left( \Delta^2 (q_1\cdot q_2) - 2 (q_1\cdot \Delta) (q_2 \cdot \Delta) \right) g^{\mu \nu} - \Delta^2 q_1^\nu q_2^\mu - 2(q_1 \cdot q_2) \Delta^\mu \Delta^\nu + 2 (q_2 \cdot \Delta) q_1^\nu \Delta^\mu + 2 (q_1 \cdot \Delta) q_2^\mu \Delta^\nu , \nonumber \\
    K_3^{\mu \nu} &= (t-u) \Big[ \left(Q_1^2 (q_2 \cdot \Delta) - Q_2^2 (q_1 \cdot \Delta) \right) \left(g^{\mu\nu}- \frac{q_1^\nu q_2^\mu}{q_1 \cdot q_2}\right)  \nonumber \\
    &- \left(\Delta^\nu - \frac{(q_2 \cdot \Delta)q_1^\nu}{q_1 \cdot q_2}\right) \left(Q_1^2 q_2^\mu + q_1^\mu (q_1 \cdot q_2)\right) 
    + \left(\Delta^\mu - \frac{(q_1 \cdot \Delta)q_2^\mu}{q_1 \cdot q_2}\right) \left(Q_2^2 q_1^\nu + q_2^\nu (q_1 \cdot q_2)\right) \Big] \nonumber \\
    K_4^{\mu \nu} & = Q_1^2 Q_2^2 g^{\mu \nu} + Q_1^2 q_2^\mu q_2^\nu + Q_2^2 q_1^\mu q_1^\nu + q_1^\mu q_2^\nu (q_1 \cdot q_2) , \nonumber \\
    K_5^{\mu \nu} & = \left(Q_1^2 \Delta^\mu + (q_1 \cdot \Delta) q_1^\mu \right) \left(Q_2^2 \Delta^\nu + (q_2 \cdot \Delta) q_2^\nu \right) , \nonumber 
\end{align}
with $\Delta \equiv p_1 - p_2 = P - 2p$, with $q^\mu_i,\,p^\mu_i$ being the photon and pion four-momenta, respectively.

 In order to then project Eq.~\eqref{eq:genTdecomp}, we first expand the form factors in~\cref{eq:genTdecomp} over partial waves
\begin{align}
	h_j(s,t)=\sum_\ell (2\ell+1) h_{j,\,\ell}(s) P_\ell(\cos\theta) \, ,
	\label{eq:Fjl}
\end{align}
where $h_{j,\,\ell}(s)$ are defined to have definite angular momentum $\ell$ and $\cos\theta$ is the scattering angle of the $\pi\pi$ system in its CM frame.

Although the Lorentz decomposition above holds in general, it is convenient to write Eq.~\eqref{eq:genTdecomp} in the CM of the system, denoted by $\star$, where
\begin{align}
	q_1^\mu &=(\omega^\star,0,0,{q}^\star),
	\label{eq:q1s}
	\\
	q_2^\mu  &=(\sqrt{s}-\omega^\star,0,0,-{q}^\star),
	\label{eq:q2s}
	\\
	p_1^\mu &=(\sqrt{s}/2,p^\star\sin\theta \cos\phi,p^\star\sin\theta \sin\phi,p^\star\cos\theta),
	\label{eq:p1s}
	\\
	p_2^\mu  &=(\sqrt{s}/2,-p^\star\sin\theta \cos\phi,-p^\star\sin\theta \sin\phi,-p^\star\cos\theta) \, ,
	\label{eq:p2s}
\end{align}
where the fact that pions are on-shell implies that $p^2_1=p^2_2=m^2$, or equivalently $p^{\star\,2} = s/4 - m^2$. Using this and Eq.~\eqref{eq:Fjl}, one can integrate over the solid angle, as dictated by Eq.~\eqref{eq:PW}, to find the $0^{++}$ component of the amplitude. 
Following this procedure and writing the final expression in terms of Lorentz tensors, one finds that the amplitude can be decomposed as 
\begin{align}
	\label{eq:scalarTdecomp}
    \Tc^{\mu\nu}(P;q_1) = \hat{h}_1(s) K_1^{\mu \nu}+\hat{h}_2(s) \hat{K}_2^{\mu \nu},
\end{align}
where the tensor $\hat{K}^{\mu\nu}_2$ is
\begin{align}
    \label{eq:khat2}
    \hat{K}^{\mu\nu}_2=(q1\cdot q2) q_1^\mu q_2^\nu+Q_1^2 q_2^\mu q_2^\nu+Q_2^2 q_1^\mu q_1^\nu+\frac{Q_1^2 Q_2^2 q_2^\mu q_1^\nu}{q_1 \cdot q_2}.
\end{align}
The $s$-dependent form factors, $\hat{h}_1$ and $\hat{h}_2$, can be written in term of the $h_{j,\ell}$ as

%%%%%%%%%%%%%%%%%%
%	figure
%%%%%%%%%%%%%%%%%%
\begin{figure}[t]
	\begin{center}
	\includegraphics[width=1\textwidth]{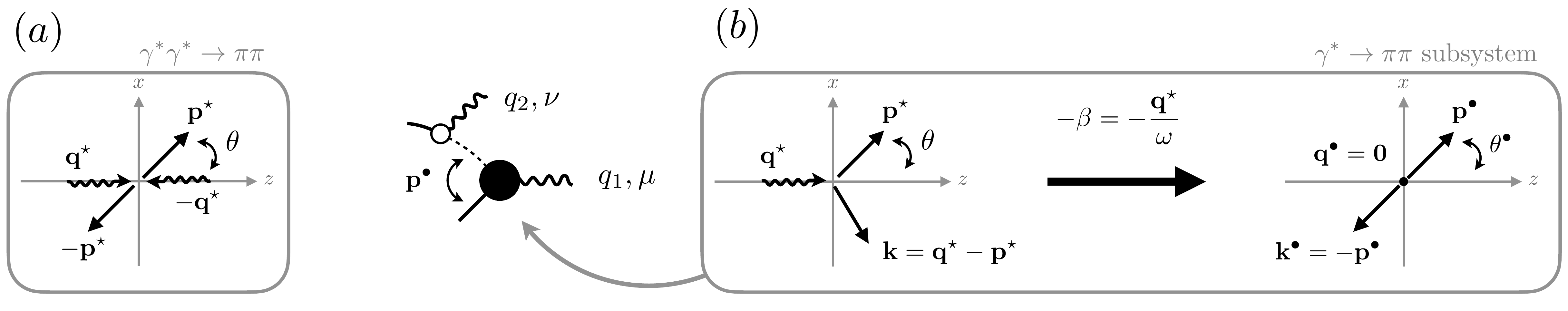}
	\caption{Kinematics for the (a) $\gamma^{\star}\gamma^{\star}\to \pi\pi$ system evaluated in the final CM frame ($\mathbf{P} = \mathbf{0}$ with $\star$ indicators), and (b) the $\gamma^{\star} \to \pi\pi$ sub-process evaluated both in the final state CM frame (left) and boosted to the intermediate state CM frame ($\mathbf{q}_1 = \mathbf{0}$ with $\bullet$ indicators).
	}
	\label{fig:frames}
	\end{center}
\end{figure}
%%%%%%%%%%%%%%%%%%
%	figure
%%%%%%%%%%%%%%%%%%

%
\begin{subequations}
	\begin{align}
	\label{eq:scalar_functions}
		\hat{h}_1 & = h_{1,0} - \frac{Q_1^2 Q_2^2 }{q_1\cdot q_2} \, h_{4,0} + \frac{8 p^{\star\,2} Q_1^2 Q_2^2}{6(q_1\cdot q_2)} \, \left( h_{5,0} - h_{5,2} \right) \, \nn \\[5pt]
		& \qquad - \frac{2p^{\star\,2}}{3s(q_1\cdot q_2)} \, \left[ \lambda(s,-Q_1^2,-Q_2^2) - 2s (q_1\cdot q_2)\right] \, \left( h_{2,0} + 2 h_{2,2} \right) \nn \\[5pt]
		& \qquad - \frac{2p^{\star\,2}}{3s(q_1\cdot q_2)} \, (Q_1^2 + Q_2^2) \, \lambda(s,-Q_1^2,-Q_2^2) \, \left( h_{3,0} + 2 h_{3,2} \right) \, , \\[5pt]
		\hat{h}_2 & = h_{4,0} - \frac{8p^{\star\,2}}{3s} \, \left( h_{2,0} + 2 h_{2,2} \right) + \frac{8p^{\star\,2}}{3} \left(  h_{3,0} + 2 h_{3,2}  \right) \, \nn \\[5pt]
		& \qquad + \frac{2p^{\star\,2}}{3s} \, \left( Q_1^2 + Q_2^2 - s\right) \, \left(  h_{5,0} + 2 h_{5,2} \right) \, \nn \\[5pt]
		& \qquad + \frac{16 p^{\star\,2}}{\lambda(s,-Q_1^2,-Q_2^2)} \, \left[ 2(q_1\cdot q_2) h_{2,2} - Q_1^2 Q_2^2 \, h_{5,2} \right] \, ,
	\end{align}
\end{subequations}
where $2 (q_1 \cdot q_2) = s + Q_1^2 + Q_2^2$, and $\lambda(a,b,c) = a^2 + b^2 + c^2 - 2(ab + bc + ca)$ is the K\"allén triangle function.

Hence, as was expected, if the isoscalar-scalar wave is dominant, the \ggpipi amplitude can be parameterized simply by two form factors. However, $\Tc^{\mu\nu}_{\df}$ is the quantity that may be most readily accessible from lattice QCD, not $\Tc^{\mu\nu}$. Nonetheless, as explained in~\cref{eq:Tmunu}, we can relate the two through the pion-pole pieces $\sum \left\{iw^{\mu}_{\mathrm{on}}\,iD\,i\overline{\Hc}^{\nu}\right\}$. These contributions contain explicit angular dependence on the intermediate state produced by the exchange of an off-shell pion either in the $t$ or $u$ channels. This process must be projected onto the S-wave final state to be related to the formulae above. There are a total of four possible pion-pole contributions. The first two account for $q_1,\,\mu$ on the initial state, where $q_2,\,\nu$ can be inserted in either final-state pion line, the other two include the permutation $\left[q_1,\mu\leftrightarrow q_2,\nu\right]$. The first two contributions are explicitly written in Eq.~\eqref{eq:T_poles}. 

In the following we compute the explicit expression for one of the terms. This contribution, depicted in Fig.~\ref{fig:frames}, is associated with the virtual photon with momentum $q_1$ and Lorentz indice $\mu$ produces an on-shell pion with momentum $p_1$ and an off-shell pion with momentum $k=q_1-p_1$. It is this off-shell pion that subsequently couples to the other virtual photon, with momentum $q_2$ and Lorentz index $\nu$. For simplicity, we omit the more explicit notation used in Eq.~\eqref{eq:T_poles} labelling the charge associated with $w^{\nu}_{\mathrm{on}}$ matrix element. More precisely
\begin{align}
    i\Tc^{\mu \nu}_p &=iw^{\nu}_{\mathrm{on}}(p_2,p_2-q_2)iD(p_2-q_2)i\overline{\Hc}^{\mu}(\mathbf{p}_2-\mathbf{q}_2;q_1) \, ,
\end{align}
where $w^{\nu}_{\mathrm{on}}$ and $D$ were previously defined in Eqs.~\eqref{eq:won} and \eqref{eq:D_bareprop}, and $\overline{\Hc}^{\mu}$ is defined as
\begin{align}
	\label{eq:pion_pole_pieces}
    i\overline{\Hc}^{\mu}(\mathbf{k};q_1) &= \sqrt{4\pi}\sum_{ m_J=-1,0,1} \left(\frac{k^\bullet}{k_{\mathrm{on}}^\bullet}\right) Y_{1 m_J}(\hat{\mathbf{k}}^\bullet) \, i\Hc^\mu_{1 m_J}(q_1) \, .
\end{align}
As previously discussed, the difference between $\overline{\Hc}^{\mu}$ and ${\Hc}^{\mu}$ is due to barrier factors defined in the CM frame of the two-particle subsystem coupling to the current depicted here by the $\bullet$ symbol. Note that, in this formula,  $k_{\mathrm{on}}^\bullet$ represents the on-shell two-particle momentum, whereas $k^\bullet$ is the momentum of the off-shell particle. In general, this frame will differ from the CM $\pi\pi$ frame, defined in defined by Eqs.~\eqref{eq:q1s}-\eqref{eq:p2s}.  

It is worth noting that we have chosen $w^{\mu}_{\mathrm{on}}(p_2,p_1)=(p_1+p_2)^{\mu} f(Q^2_2)$ for both on- and off-shell pions. In this sense $w^{\mu}_{\mathrm{on}}$ does not respect the Ward-Takahashi identities unless both pions are on-shell. On the contrary, $\overline{\Hc}^{\nu}$ is described in terms of the on-shell pieces $\Hc^\mu_{J m_J}(q_1)$. As a result $q_{\mu}\overline{\Hc}^{\mu}=0$ for both on- and off-shell pions. This is in contrast to customary calculations where the long-range contribution is chosen so that $\overline{\Hc}^{\mu}(p_2;\vec{p_1}^\star;q_1)=(p_1-p_2)^{\mu} f(-Q^2_1)$ even for off-shell pions.  

In order to explicit partial wave project the pole contribution, which will be done in the CM frame of the $\pi\pi$ system, we first write the $k^{\bullet \mu}$ momenta in terms of $k^{\star\,\mu}$ using a standard Lorentz boost. In particular, $k^{\bullet\,\mu}=\left[\Lambda_{-\beta}\right]^\mu_{\,\,\,\nu}\,k^{\star \,\nu}$, where the boost vector points along the $\hat{z}$ axis and has magnitude $\beta = q^\star/\omega$. Additionally, we make use of the relation between spherical harmonics and Cartesian coordinates to write everything in terms of the $\pi\pi$ CM coordinates. For example,
\begin{align}
    \label{eq:y10}
   \sqrt{\frac{4\pi}{3}} k^\bullet Y_{1,0}(\theta^\bullet,\phi^\bullet)=k^{\bullet\,z}=\left[\Lambda_{-\beta}\right]^z_{\,\,\,\nu}\,k^{\star \,\nu} \, .
\end{align}
In this way, we established a simple relation between the $\theta^\bullet,\phi^\bullet $ angles and $k^{\star \,\nu}$, which contains the explicit $\theta,\phi$ dependence on its frame.

Once the spherical harmonics have been rewritten following this approach, the next step is to perform the partial-wave projection onto the scalar state by means of the formula 
\begin{align}
    i\Tc^{\mu \nu}_{p,\,0^{++}}=\frac{1}{4\pi} \,  
    \int_{-1}^{1} \!\diff{\cos\theta}
    \,\int_{0}^{2\pi} \! \diff{\phi}
    \,   \, iw^{\nu}_{\mathrm{on}}(p_2,p_2-q_2)iD(p_2-q_2)i\overline{\Hc}^{\mu}(\mathbf{p}_2-\mathbf{q}_2;q_1) \, ,
    \label{eq:scalar_projection}
\end{align}

It is advantageous to decompose both vector expressions in the right-hand side by polarization vectors. These provide a compact expression for the current insertion term. After performing the integrations we can regroup all terms as 
\begin{align}
    \label{eq:pion_pole_projected}
     \Tc^{\mu \nu}_{p,\,0^{++}}
     =f(Q_2^2) f(-Q_1^2)\,
     \Big[ \, 
     c_0\, \epsilon^{\mu*}_0(q_1)\epsilon^{\nu*}_0(0)
     +c_1\, (\epsilon^{\mu *}_+\epsilon^{\nu*}_- +\epsilon^{\mu *}_-\epsilon^{\nu*}_+)
     +c_L \, \epsilon^{\mu *}_0(q_1)\epsilon^{\nu*}_L
     \, \Big] \, ,
\end{align}
where $k^\bullet $ is the intermediate off-shell pion momentum on the CM frame of the $q_1$ photon and
\begin{align}
	\epsilon^{\mu}_0(q)&=\frac{1}{\sqrt{-Q^2_1}}\left(q^{\star},0,0,\omega^{\star}\right), \nn \\
	\epsilon^{\mu}_L&=\left(1,0,0,0\right), \nn \\
	\epsilon^{\mu}_\pm&=\frac{\mp 1}{\sqrt{2}}\left(0,1,\pm i,0\right).
\end{align}
Finally, the scalar functions are given by
\begin{align}
    \label{eq:F'scalars}
    c_0 &= \frac{ \left(3 \sqrt{s} q^{\star\,2}+4 \, p^{\star\,2} \,\, \omega^\star \right) \, \mathcal{Q}_0(z) -6 p^\star  q^\star \left(\sqrt{s}+\,\omega^\star \right) \mathcal{Q}_1(z) + 8 p^{\star\,2}\,\omega^\star
    \mathcal{Q}_2(z) }{6 \, p^{\star}q^\star \, \sqrt{-Q_1^2} } \, , \nn \\[5pt]
    c_1 &= -\frac{2p^{\star} \left[\mathcal{Q}_0(z)-\mathcal{Q}_2(z)\right]}{3 q^\star} \, , \nn \\[5pt]
    c_L &= \frac{\,\omega^\star \left[\sqrt{s} q^\star \mathcal{Q}_0(z)  - 2 p^\star\,\omega^\star
   \mathcal{Q}_1(z)\right]}{2 \, p^{\star}q^\star \, \sqrt{-Q_1^2} } \, ,
\end{align}
%}
where $\mathcal{Q}_{\ell}(z)$ are the Legendre functions of the second kind, which have an argument $z=\left(Q_1^2+\sqrt{s}\omega^\star \right)/(2 p^\star q^\star)$, where $ q^\star = \lambda^{1/2}(s,-Q_1^2,-Q_2^2) / (2\sqrt{s} )$. As a final remark, it is worth noting that the term proportional to $\epsilon^{\nu}_L$ is the one that explicitly violates the Ward-Takahashi identities.

%%%%%%%%%%%%%%%%%%%%%%%%%%%%%%%%%%%%
%	App :: Finite-Volume Functions
%%%%%%%%%%%%%%%%%%%%%%%%%%%%%%%%%%%%
\section{Finite-volume functions: $F$, $G$, and $G^\mu$}
\label{app:FGL}

In this appendix, we give the exact forms of the finite-volume geometric functions described in the text. The first function is $F(P,L)$ defined implicitly in Eq.~\eqref{eq:on_shell_loop} and depicted diagrammatically in Fig.~\ref{fig:Loop}. This has been well described in the literature, and here we follow the definition first introduced in Ref.~\cite{Kim:2005gf}. In general, this is a matrix over open channels and partial waves. Assuming a single open channel of identical scalar particles of mass $m$, this is a matrix in angular momentum space with components given by 
\begin{align}
	\label{eq:FV_F}
    F_{\ell m_\ell ;\ell' m'_\ell}(P,L) \equiv \xi \left[\frac{1}{L^3}\sum_{\mathbf{k}} - \int\! \frac{\diff^3 \textbf{k}}{(2\pi)^3}\right] \frac{1}{2\omega_{\textbf{k}}} \,\Yc_{\ell m_\ell }^*(\mathbf{k}^{\star},P) \, D\left(P-k\right) \, \Yc_{\ell' m'_\ell}(\mathbf{k}^{\star},P)\Big \rvert_{k_0 =\omega_{\mathbf{k}}} \, ,
\end{align}
where the sum of $\mathbf{k}$ is over the quantized momenta $\mathbf{k} = 2\pi \mathbf{n} / L$ for $\mathbf{n} \in \mathbb{Z}^3$, $\omega_{\mathbf{k}} = \sqrt{m^2 + \mathbf{k}^2}$ is the on-shell energy, $D$ is the pole piece of the single particle propagator defined in Eq.~\eqref{eq:D_bareprop}, and $\mathcal{Y}_{\ell m_\ell }$ are modified spherical harmonics,
\begin{equation}
    \mathcal{Y}_{\ell m_\ell }(\textbf{k}^{\star},P) \equiv \sqrt{4\pi} \, Y_{\ell m_\ell }(\hat{\mathbf{k}}^{\star})\left( \frac{k^{\star}}{q^{\star}}\right)^{\ell} \, ,
\end{equation}
with $q^{\star} = \sqrt{s/4 - m^2}$ being the on-shell relative momentum in the CM frame. Note that angular momentum is not a good quantum number in a cubic volume. This is manifested here by the fact that $F$ is a non-diagonal matrix in $\ell m_{\ell};\ell' m_{\ell}'$ space, which can be seen by the sum term in Eq.~\eqref{eq:FV_F} which in general does not vanish for different angular momenta.

The other geometric function needed is $G^{[\mu]}(P_f,P_i,L)$, which was written in Eq.~\eqref{eq:Gmu_tot} in terms of two functions, $G(P_f,P_i,L)$ and $G^{\mu}(P_f,P_i,L)$. These class of functions were studied in detail in Ref.~\cite{Baroni:2018iau}, and are defined for equal mass scalar particles as
\begin{align}
	\label{eq:FV_G}
    G_{\ell m_\ell ;\ell' m'_\ell}(P_f,P_i,L) & = \left[\frac{1}{L^3}\sum_{\mathbf{k}} - \int \! \frac{\diff^3 \mathbf{k}}{(2\pi)^3}\right] \frac{1}{2\omega_{\mathbf{k}}} \, \Yc_{\ell m_\ell }^{*}(\mathbf{k}_f^{\star},P_f) \, D(k_f) D(k_i) \, \mathcal{Y}_{\ell' m'_\ell}(\textbf{k}_i^{\star},P_i)\Big \rvert_{k_0 =\omega_{\mathbf{k}}} \, , \nn \\[5pt]
    G^{\mu}_{\ell m_\ell ;\ell' m'_\ell}(P_f,P_i,L) & = \left[\frac{1}{L^3}\sum_{\mathbf{k}} - \int \! \frac{\diff^3 \mathbf{k}}{(2\pi)^3}\right] \frac{k^\mu}{2\omega_{\mathbf{k}}}  \, \Yc_{\ell m_\ell }^{*}(\mathbf{k}_f^{\star},P_f) \, D(k_f) D(k_i)  \,  \Yc_{\ell' m'_\ell}(\mathbf{k}_i^{\star},P_i)\Big \rvert_{k_0 =\omega_{\mathbf{k}}} \, ,
\end{align}
where $k_i = P_f - k$ and $k_i = P_i - k$, and $\mathbf{k}_i^{\star}$ ($\mathbf{k}_f^{\star}$) is the summation/integration momentum $\mathbf{k}$ evaluated in the CM frame of the initial (final) state. We point the reader to Ref.~\cite{Baroni:2018iau} for further details on efficient numerical techniques for evaluating this class of functions.

%%%%%%%%%%%%%%%%%%%%%%%%%%%%%%%%%%%%
%	App :: Bound state limit
%%%%%%%%%%%%%%%%%%%%%%%%%%%%%%%%%%%%

\section{The deeply bound state limit}
\label{app:deep_bs}

Here we consider a simple limit to check the normalization appearing in the main result, Eq.~\eqref{eq:Main_eq}. In particular, we assume the presence of a deeply bound state with mass $m_B$ and four-momentum $P_B = (E_B, \mathbf{P})$. This could be, for example, the $\sigma$ for unphysically heavy quark masses, where lattice QCD calculations observe it to be bound (see, for example, Ref.~\cite{Briceno:2016mjc}). We will choose the volume such that $\kappa_B L\gg 1$, where $\kappa_B$ is the binding momentum of the two-particle state. This allows us to ignore exponentially suppressed effects associated with the size of the bound state that scale as $\mathcal{O}(e^{-\kappa_B L})$. Furthermore, we will restrict the momenta of the currents such that no intermediate states can go on shell. One simple example is to fix $q_1=P_B/2$.

This was the limit previously considered in Ref.~\cite{Briceno:2019nns}. There it was shown that both $F$ and $G^{[\mu]}$ scale as $\mathcal{O}(e^{-\kappa_B L})$. Ignoring such terms allows us to simplify Eq.~\eqref{eq:Main_eq} substantially. In particular, $\Delta \Tc^{\mu\nu}_{L,\df}$, given in Eq.~\eqref{eq:deltaTLdf}, can be approximated to be equal to zero. Furthermore, because of the kinematics chosen, no intermediate states can go on shell. As a result, there is no need to subtract any terms from the time-dependent correlation function.

With these two observations, Eq.~\eqref{eq:Main_eq} simplifies down to
\begin{align}
	\label{eq:Main_eq_bs}
	\frac{r_B}{\sqrt{2 E_B L^3}}
	\cdot
	\Tc^{\mu\nu}_{\df}(P;q_1)
	&=
	\int \diff \tau \, e^{\omega\tau} \,
	M_L^{\mu\nu}(\tau,P; \textbf{q}_1) \, ,
\end{align}

where  we have ignored all partial waves other than the one coupling to the bound state, making $\rrvec$ in a multiplicative factor denoted $r_B$. Reference~\cite{Sherman:2022tco} showed that for such a case, $\Tc^{\mu\nu}_{\df}$ must have a pole associated with the bound state given by
\begin{align}
	\label{eq:T_pole}
	\lim_{s \to m_B^2}(s-m_B^2) \, \Tc^{\mu\nu}_{\df}(P;q_1) = - g\, F^{\mu\nu}(P_B;q_1) \, ,
\end{align}
where $s = P^2$, the momentum of the bound state where $P_B^2 = m_B^2$, $F^{\mu\nu}(P;q_1)$ is the two-current form factor for the bound state, and $g$ is the coupling to the bound state to the two-particle scattering system. The latter is given from the residue of the two-particle scattering amplitude at the bound state pole, 
\begin{align}
	\label{eq:M_pole}
	\lim_{s \to m_B^2}(s-m_B^2) \, \Mc(P^2) = - g^2 \, .
\end{align}

Given that the right-hand side of Eq.~\eqref{eq:Main_eq_bs} is by definition finite, this means that $r_B$ must vanish in this limit as $s-m_B^2$. Here we reproduce the calculation showing that this is indeed the case, and we show that the resulting normalization is the correct one for such bound state. We proceed from the definition of $r_B$, which follows from Eqs.~\eqref{eq:F2L_pole},~\eqref{eq:Rc_vf}, and~\eqref{eq:f2l}. Because this vanishes as $s-m_B^2$, we will calculate the factor multiplying $s-m_B^2$, 
\begin{align}
	\frac{r_B }{s-m_B^2}
	&=
	\frac{\sqrt{2E_B}}{s-m_B^2}\sqrt{\frac{E-E_B}{F^{-1}(P,L)+\Mc(P^2)}}\bigg|_{E = E_B},
	\nn\\[5pt]
	&=
	\frac{\sqrt{-2E_B}}{g^2}
	\sqrt{\frac{E-E_B}{F(P,L)+\Mc^{-1}(P^2)}}\bigg|_{E = E_B},
	\nn\\[5pt]
	&\approx
	\frac{1}{g^2}
	\sqrt{-\frac{E^2-E_B^2}{\Mc^{-1}(P^2)}}\bigg|_{E = E_B},
	\nn\\[5pt]
	&=
	\frac{1}{g} \, ,
\end{align}

where in the second line we use the fact that at the pole $F=-\Mc^{-1}$, and also used the behavior of $\Mc$ near its pole, Eq.~\eqref{eq:M_pole}. In the third line, we ignore higher order terms proportional to $\mathcal{O}(e^{-\kappa_B L})$. Putting this together with Eq.~\eqref{eq:T_pole}, we obtain
\begin{align}
	\label{eq:Main_eq_bs_2}
	\frac{-F^{\mu\nu}(P_B;q_1)}{\sqrt{2 E_B L^3}}
	&=
	\int \diff \tau \, e^{\omega\tau} \, 
	M_L^{\mu\nu}(\tau,P_B; \textbf{q}_1).
\end{align}
The overall sign is not physical and can be ignored. The remaining factor of $\sqrt{2 E_B L^3}$ is exactly what is needed to fix the normalization of the finite-volume state that was assumed to be normalized to $1$. Accordingly, we conclude that the main result, Eq.~\eqref{eq:Main_eq}, reproduces the expected behavior for a long-range matrix element involving a bound state in the final state.

%%%%%%%%%%%%%%%%%%%%%%%%%%%%%%%%%%%%
%	Bibliography
%%%%%%%%%%%%%%%%%%%%%%%%%%%%%%%%%%%%
\bibliographystyle{apsrev4-1}
\bibliography{bibi.bib}

%%%%%%%%%%%%%%%%%%%%%%%%%%%%%%%%%%%%
%	End document
%%%%%%%%%%%%%%%%%%%%%%%%%%%%%%%%%%%%
\end{document}